\documentclass{nature}
\usepackage{amssymb}
\usepackage{aas_macros}
\usepackage{graphicx}
\usepackage{pdflscape}
\usepackage{hyperref}
\usepackage{newtxtext,newtxmath}
\usepackage{textcomp}
\usepackage{threeparttable}
\usepackage{xcolor}
\usepackage{ulem}
\usepackage{lineno}
\usepackage{pdfpages}
\usepackage[caption = false]{subfig}
\makeatletter
\usepackage[symbol]{footmisc}

\usepackage{longtable}

\usepackage[font={small,sf}, singlelinecheck=false]{caption}
\let\saved@includegraphics\includegraphics
\AtBeginDocument{\let\includegraphics\saved@includegraphics}
\renewenvironment*{figure}{\@float{figure}}{\end@float}
\makeatother

\title{Unveiling Dominant Toroidal Magnetic Fields in a Protostellar Outflow}

\author{
T.-C.\ Ching$^{1}$\thanks{Email:tching@nrao.edu, Tao-Chung Ching \textcolor{red}{was} a Jansky Fellow of the National Radio Astronomy Observatory, \href{https://orcid.org/0000-0001-8516-2532}{orcid.org/0000-0001-8516-2532}},
Z.-Y.\ Li$^{2}$,
Q.\ Zhang$^{3}$,
J.\ M.\ Girart$^{4,5}$,
S.-P.\ Lai$^{6}$,
C.-F.\ Lee$^{7}$,
D.\ Li$^{8,9,10}$,
R.\ Rao$^{3}$,
E.\ Momjian$^{1}$}

\begin{document}
%\linenumbers
\maketitle

\begin{affiliations}
\item National Radio Astronomy Observatory, 1011 Lopezville Road, Socorro, NM 87801, USA
\item Astronomy Department, University of Virginia, Charlottesville, VA 22904, USA
\item Center for Astrophysics $|$ Harvard $\&$ Smithsonian, 60 Garden Street, Cambridge, MA 02138, USA
\item Institut de Ciències de I’Espai (ICE-CSIC), Campus UAB, Carrer de Can Magrans s/n, E-08193 Barcelona, Catalonia, Spain
\item Institut d’Estudis Espacials de Catalunya (IEEC), E-08860 Castelldefels, Catalonia, Spain
\item Institute of Astronomy and Department of Physics, National Tsing Hua University, Hsinchu 30013, Taiwan
\item Academia Sinica Institute of Astronomy and Astrophysics, No.\ 1, Sec.\ 4, Roosevelt Road, Taipei 10617, Taiwan
\item Department of Astronomy, Tsinghua University, Beijing 100084, People’s Republic of China
\item National Astronomical Observatories, Chinese Academy of Sciences, Beijing 100101, People’s Republic of China
\item Research Center for Astronomical Computing, Zhejiang Laboratory, Hangzhou 311100, People’s Republic of China
\end{affiliations}

\begin{abstract}
Magnetic fields play a fundamental role in the formation of protostellar winds. In the magneto-centrifugal models, poloidal magnetic fields launch winds from accretion disks, and fast-rotating gas twists the fields into toroidal geometry {that collimates and accelerates winds through magnetic hoop stress.} However, toroidal fields {in protostellar} winds remain observationally unresolved. Here we report polarization observations of carbon monoxide emission toward the NGC1333 IRAS 4A protostellar outflow. The inferred magnetic fields are perpendicular to the outflow axis and aligned with the rotational structure of the outflow, indicating toroidal fields {with strengths of a few milligauss, sufficient to collimate and accelerate the outflow} at several hundred astronomical units from the protostar. A linear correlation is found between the curl of plane-of-the-sky magnetic field and the line-of-sight {electric current density.
Our analysis provides better constraints on ion–electron drift velocity in protostellar outflows and supports rotating outflows driven by the magneto-centrifugal mechanism.}
%Magnetic fields play a fundamental role in the formation of protostellar winds. In the magneto-centrifugal models, poloidal fields are required to launch winds from accretion disks, and fast-rotating gas twists the fields into toroidal geometry. Due to the hoop stress of the toroidal fields, winds are self-collimated and accelerated to a high velocity. However, toroidal fields of winds remain observationally unresolved. Here we report polarization observations of carbon monoxide (CO) emission toward the NGC1333 IRAS 4A protostellar outflow. The inferred magnetic fields are perpendicular to the outflow axis and aligned with the rotational structure of outflow, indicating toroidal fields wrapping around the outflow at several hundred astronomical units from the protostar. The field strengths of a few milligauss are strong enough to collimate and accelerate the outflow. Based on the Amp\`ere's law, a linear correlation is found between the curl of the plane-of-the-sky magnetic field inferred from the CO polarization and the line-of-sight current inferred from the CO total emission. Our method of analyzing molecular line polarization observations provides better constraints on key physical parameters, such as the ion-electron drift velocity, for outflow simulations and confirms the classical picture of rotating outflows driven by the magneto-centrifugal mechanism. 
\end{abstract}

\section{Introduction}
Collimated outflows are a common feature of material accretion in various astrophysical systems from active galactic nuclei to young stellar objects. While these systems differ vastly in physical scales, the launching of outflows is widely attributed to centrifugal driving by hydromagnetic flows of an accretion disk surrounding a central object\cite{1982BP}.
%are required to launch winds from accretion disks, and the fast-rotating gas of wind carries angular momentum off the disk, twisting the magnetic fields into toroidal geometry. 
In the magneto-centrifugal models of protostellar winds\cite{2000Shu,2007Pudritz}, {large-scale magnetic field lines guide centrifugally accelerated gas away from the accretion disk, launching initially sub-Alfv\'enic wind co-rotating with poloidal fields near the disk. At larger distances, the wind is accelerated to a super-Alv\'enic speed, where the magnetic field becomes predominantly toroidal.}
Due to the hoop stress of the toroidal fields, winds are self-collimated and accelerated to form high-velocity jets, and the ambient material swept up by the jets forms low-velocity outflows with wider opening angles.
However, the self-collimation in the magneto-centrifugal model is challenged by the kink instability induced by toroidal fields, which can potentially disrupt the wind\cite{2008Moll}.
Alternative mechanisms for outflow collimation, such as external thermal pressure\cite{1982K} or large-scale poloidal magnetic fields\cite{1997Spruit}, have been explored, and laboratory experiments have yielded contradictory evidence on whether toroidal or poloidal magnetic fields primarily collimate outflows\cite{2014Albertazzi,2024Lei}.%, leaving the dominant magnetic field structure of astronomical outflows unresolved.

Despite the central prediction of the magneto-centrifugal model that the magnetic fields of jets and outflows {at large distances from disk} are predominantly toroidal, it has rarely been observed before. 
Near-infrared circular polarimetry toward the Herbig–Haro (HH) 135–136 object suggests that the magnetic fields along the bipolar outflow are helical\cite{2007Chrysostomou}. % at a distance of a few thousand astronomical units (au) from the central protostar. 
Observations of polarized synchrotron emission in the HH 80-81 object also suggest a helical magnetic field structure of the jet\cite{2010Carrasco} which can be interpreted with a dominant but unresolved toroidal component in the center and a poloidal component at the edge\cite{2025Rodriguez}.
{In the case of IRAS 21078+5211, the three-dimensional velocity field inferred from water masers reveals a rotating helical structure, which is consistent with a magneto-centrifugal disk wind model\cite{2022Moscadelli} that includes both poloidal and toroidal magnetic field components; the latter can contribute to both outflow collimation and acceleration\cite{2023OKb}.}
Additionally, the linear polarization observations of silicon monoxide (SiO) emission in the HH 211 jet imply that the magnetic fields could be either mainly poloidal or mainly toroidal\cite{2018Lee}.
%Because the previous studies of magnetized jets and outflows rely on radiation integrated along the line of sight, the line-of-sight information of magnetic fields is averaged out. 
{Because most previous polarization studies lacked spectral resolution to resolve magnetic field structures in velocity along the line of sight, the line-of-sight  variations of magnetic fields within jets and outflows were averaged out.}
%{Because most previous studies rely on polarized radiation integrated over the line-of-sight depth of jets and outflows, the line-of-sight variation of magnetic fields within jets and outflows is averaged out.}
As a result, interpreting the data requires complex modeling and multiple assumptions, making the inferred magnetic field structure elusive.

We performed polarization observations of the carbon monoxide {(CO)} $J = 2–1$ rotational transition toward the protostellar core NGC1333 IRAS 4A using the Atacama Large Millimeter/submillimeter Array (ALMA).
IRAS 4A is a well-studied low-mass core in Class 0 phase, harboring two protostars IRAS 4A1 and 4A2, located in the star-forming region NGC 1333 of the Perseus molecular cloud complex\cite{1991Sandell,2000Looney} at a distance of 293 $\pm$ 22 parsecs (pc) from us\cite{2018Ortiz}.
The magnetic field of IRAS 4A traced by dust polarization exhibits an hourglass morphology at a scale of a few hundred astronomical units (au)\cite{2006Girart}, consistent with the theoretical prediction of a collapsing core {in the presence of ordered magnetic fields}\cite{2008Goncalves,2011Frau}.
Highly collimated high-velocity {SiO} jets, one from each of the IRAS 4A1 and 4A2, trace the on-axis density enhancement of protostellar winds\cite{2006Choi}, and low-velocity outflow cavities revealed in multiple molecular emission tracing shells of ambient gas swept-up by the winds\cite{1999Girart,2019Su,2024Chahine}. %cite Chuang 2021 ?

The CO linear polarization arises from the Goldreich-Klafis effect (GK effect)\cite{1981GK}, which predicts that 
%an unequal population of the magnetic sublevels of a molecular rotational transition, combined with anisotropic optical depth, produces net linearly polarized emission.
{when the magnetic precession rate exceeds collisional and radiative rates, anisotropies in optical depth, radiation field, and velocity gradients can lead to unequal populations of magnetic sublevels in a molecular rotational transition, resulting in net linearly polarized emission.}  
The polarization orientation is either perpendicular or parallel to the projected magnetic field on the plane of the sky.
In general, the GK effect produces a maximum polarization when the velocity flow and the magnetic field are both perpendicular to the line of sight\cite{1984DW}, and therefore most of the measurements of the GK effect have been obtained toward protostellar jets and outflows lying close to the plane of the sky (e.g.\ refs.\ \cite{2003Lai,2018Lee,2020Hirota,2021Cortes,2023Barnes}).
The GK effect of the CO $J = 2–1$ and $3–2$ transitions has been mapped in the IRAS 4A outflows with the Berkeley-Illinois-Maryland association Array (BIMA)\cite{1999Girart} and the Submillimeter Array (SMA)\cite{2016Ching}, but the angular resolutions of those polarization data cannot resolve the binary and constrain whether the magnetic field structure of the outflow is dominated by toroidal or poloidal components.

\section{Results}
\subsection{CO linear polarization in the IRAS 4A outflow.}
Fig.\ \ref{fig1} shows the ALMA CO $J = 2-1$ polarization maps at four velocity channels toward IRAS 4A. 
The two bipolar outflows of 4A1 and 4A2 are well resolved with a spatial resolution of 147 au $\times$ 99 au, which is 31 times higher than those of the BIMA and SMA CO polarization observations.
The CO redshifted outflows show an initial morphology in the northwest direction from the launching protostars and become northeast oriented at a distance of about 1000 au to the north of the protostars, as unveiled in multiple outflow tracers\cite{2019Su,2024Chahine}.
The ALMA CO $J = 2-1$ polarization measurements are extended close to 4A1 with polarization angles in the northeast-southwest orientation. The ALMA polarization orientation around 4A1 is parallel to those of the BIMA CO $J = 2-1$ polarization and SMA CO $J = 3-2$ polarization. At the northern part of the outflow where the outflow direction changes from northwest to northeast, the ALMA polarization measurements become scattered.

To solve the ambiguity of a parallel or perpendicular orientation between the GK effect and the magnetic field, {comparisons between spectral-line polarization and dust polarization are considered essential in studying magnetic field through the GK effect\cite{2022Houde} and also shown to be effective in magnetohydrodynamic (MHD) simulations\cite{2022Bino}.}
Fig.\ \ref{fig2}a shows the integrated CO polarization measurements overlaid with the magnetic field inferred from the dust polarization obtained in our ALMA data.
The CO polarization orientation is generally parallel to the magnetic field orientation.
%, in agreement with our theoretical calculation of the CO GK effect in the 4A1 redshifted outflow\cite{2016Ching}.
Small deviations in the position angles between the CO polarization and the magnetic field inferred from the dust polarization arise since the CO polarization traces the magnetic field in the outflow cavity of the dusty core, slightly different from the magnetic field of the core revealed by the dust polarization.
The outflow velocity field (Fig.\ \ref{fig2}b) shows that the eastern side of the redshifted outflow has a faster line-of-sight velocity than the western side.
Fig.\ \ref{fig2}c shows the position-velocity (PV) diagram along the minor axis of the 4A1 outflow. %A linear velocity gradient of 20.5 $\pm$ 4.0 au/(km s$^{-1}$) in the PV diagram indicates a rotating outflow with a ring-like cavity structure\cite{2017Hirota}.
A linear velocity gradient of {$(4.9 \pm 1.0) \times 10^{-2}$ km s$^{-1}$ au$^{-1}$} in the PV diagram indicates a rotating outflow with a ring-like cavity structure\cite{2017Hirota}.
The parabolic features in the PV diagram along the major axis of the 4A1 outflow (Fig.\ \ref{fig2}d) are consistent with the cavity shells swept up by a wide-angle protostellar wind.
Our fittings of the parabolic features using an empirical equation of wind-driven model\cite{2000Lee} indicate that the parabolic features can be generated by one protostellar wind with two opening angles (see Methods for the fittings). 

{Modern three-dimensional simulations that incorporate gas dynamics and radiative transfer have been suggested as an important tool for resolving the 90$^\circ$ ambiguity between the GK effect and the magnetic field\cite{2020LV,2024TK}.
Here we adopt a simplified code\cite{2010YL} which has been applied to observations of GK effect\cite{2016Ching,2020Huang} to solve the radiative transfer equations following the formulation of GK effect under large velocity gradient approximation\cite{1984DW} and an external anisotropic radiation source\cite{2005Cortes}.
Despite being simplified compared to modern three-dimensional simulations, the large velocity gradient approximation provides sufficient accuracy for polarization predictions below 2\%\cite{2020LV}.
Using the physical properties of the IRAS 4A outflow, Fig.\ \ref{fig3new} presents the predicted CO $J = 2-1$ polarization for a toroidal magnetic field aligned with the outflow velocity gradient. The predicted polarization is parallel to magnetic field, consistent with the alignment between observed CO polarization and the magnetic field inferred from dust polarization (Fig.\ \ref{fig2}a).
The predicted polarization reaches a peak of 0.5\%, in agreement with the observed CO polarization percentage toward the center of the redshifted outflow.
For a poloidal magnetic field perpendicular to the velocity gradient, the predicted polarization percentage is reduced to 0.2\%, below the ALMA sensitivity of 0.3\% for extended polarized emission\cite{ALMA4}.
The polarization orientations of multiple CO transitions remain consistent in our setup using 2.7 K background radiation, without exhibiting the 90$^\circ$ polarization flips that can arise from anisotropic radiation fields\cite{2005Cortes} or sparse sampling of asymmetric radiative transfer conditions\cite{2020LV}.
This explains why the polarization orientations of the ALMA CO $J = 2-1$ and SMA CO $J = 3-2$ observations are consistent and further supports that ALMA CO $J = 2-1$ polarization traces toroidal magnetic fields.}
%Together, our calculation of the GK effect suggests that the orientation of the observed polarization is likely parallel to the magnetic eld, consistent with the comparison of the CO and dust polarization.

\subsection{Correlation between $\nabla \times B_{pos}$ and $J_z$.}
The CO linear polarization probes the plane-of-the-sky magnetic field structure, and the CO total emission with the information of CO velocity reveals the amount of gas moving along the line of sight.
The curl of the magnetic field strength $(B_x, B_y)$ inferred from the CO polarization and the electric current {density} $(J_z)$ inferred from the CO total emission thus should satisfy the Amp\`ere's law
\begin{equation}
\frac{\partial B_y}{\partial x}-\frac{\partial B_x}{\partial y} = \frac{4 \pi}{c} J_z +  \frac{1}{c} \frac{\partial E_z}{\partial t},
\end{equation}
where in the Equatorial coordinate system, $x$ is the axis of right ascension from west to east, $y$ is the axis of declination from south to north, and $z$ is the axis along the line of sight {(see Extended Data Figure 16 for illustration)}. 
To derive the plane-of-the-sky magnetic field strength $(B_{pos})$ from the CO polarization, we adopt the Davis--Chandrasekhar--Fermi equation\cite{1951Davis, 1953CF}.
The derived outflow $B_{pos}$ has a range from 0.3 mG to 6.0 mG (see Methods for the channels maps of $B_{pos}$). 
To derive $J_z$, we consider a pseudo-electric current {density} that $J_z = {\varrho v_d} = e n_i v_d$ with the current direction the same as the outflow direction, where {$\varrho$ is the volume charge density,} $n_i$ is the number density of ions, and $v_d$ is the drift velocity between ions and electrons.
Here $n_i$ equals $n_{H_2}$ times ionization {fraction} $X_e$ with $X_e \approx \sqrt{2} \times 10^{-5} n_{H_2}^{-1/2}$ (ref.\ \cite{Draine2011}).
{At a resolution of $\sim 5000$ au, the $X_e$ in the envelope of IRAS 4A is about 10$^{-6}$ at $n_{H_2} \sim 10^4$ cm$^{-3}$, about an order of magnitude higher than the canonical interstellar value, likely due to protostellar cosmic-ray ionization\cite{2024Pineda}.
Since the cosmic-ray ionization rate is found to be approximately constant within $\sim$ 15000 au from the young stellar objects in the NGC 1333 cloud\cite{2024Pineda}, protostellar cosmic rays primarily modify the normalization of $X_e$ rather than its density dependence, and the commonly adopted scaling $X_e \propto n_{H_2}^{-1/2}$ remains a reasonable first-order approximation.}
%Higher-resolution observations of another Class 0 protostar further indicate that $X_e$ varies from a few $\times10^{-6}$ at $n_{H_2} \sim 10^4$ cm$^{-3}$ to a peak value of $9 \times 10^{-6}$ at $n_{H_2} > 10^6$ cm$^{-3}$ (ref.\ \cite{2023Cabedo}).
Hence, Eq.\ 1 can be rewritten as
%\frac{\partial B_y}{\partial x}-\frac{\partial B_x}{\partial y} = (\frac{4 \pi e}{c} \sqrt{2} \times 10^{-5} n_{H_2}^{1/2}) v_d +  \frac{1}{c} \frac{\partial E_z}{\partial t}.
\begin{equation}
\frac{\partial B_y}{\partial x}-\frac{\partial B_x}{\partial y} = (\frac{4 \pi}{c} {\varrho}) v_d +  \frac{1}{c} \frac{\partial E_z}{\partial t}
\end{equation}
{with $\varrho = e \sqrt{2} \times 10^{-5} n_{H_2}^{1/2}$.}
Under the optically thin condition, the outflow column density can be derived from the CO total emission\cite{2015MS}. 
Since the wind-driven model fitting in Fig.\ \ref{fig2}d gives the line-of-sight depth of the 4A1 redshifted outflow, we obtain $n_{H_2}$ in a range from $3.6 \times 10^4$ cm$^{-3}$ to $6.7 \times 10^5$ cm$^{-3}$ of the outflow (see Methods for the channels maps of $n_{H_2}$). 

Fig.\ \ref{fig3} shows the correlation between the $\frac{\partial B_y}{\partial x}-\frac{\partial B_x}{\partial y}$ derived from the CO linear polarization and the $\frac{4 \pi}{c} {\varrho}$ derived from the CO total emission. 
A linear fitting of Eq.\ 2 gives a {slope} $v_d = (5.9 \pm 1.8) \times 10^{2}$ cm s$^{-1}$ and an intercept $\frac{1}{c} \frac{\partial E_z}{\partial t} = (-8.5 \pm 2.9) \times 10^{-19}$ cm$^{-3/2} $ g$^{1/2}$ s$^{-1}$.
The theoretical drift velocity between ions and neutrals in the density range of the 4A1 redshifted outflow is between $4 \times 10^3$ cm s$^{-1}$ to 2 km s$^{-1}$ (ref.\ \cite{Draine2011}), and observations give an upper limit of 0.3 km s$^{-1}$ at a distance of 100 au from a Class 0 protostar\cite{2018Yen}. 
%Our fitted $v_d$ of a few m s$^{-1}$ satisfies the observational constrain but an order of magnitude smaller than the theoretical prediction.
Our result thus suggests that $v_d$ is likely at least an order of magnitude slower than the drift velocity between ions and neutrals.
If we assume that $v_d$ is proportional to the outflow velocity, we find a weak correlation with the fitting goodness of the slope less than 1.8$\sigma$ (see Methods). 
Compared to the 3.3$\sigma$ fitting goodness when assuming a constant $v_d$, $v_d$ is less likely to be proportional to the outflow velocity.
%The value of the intercept is small as expected of a weak electric field and a slow time variation in the interstellar medium.
Since the fitting goodness of the intersection is slightly less than 3$\sigma$ {and the intercept may be systematically biased in the estimation of $B_{pos}$ (see Methods), the intersection is less physically meaningful than the slope.}

\section{Discussion}
The right-hand rule in the Amp\`ere's law requires a positive slope, and hence the sign of the slope in Fig.\ \ref{fig3} provides a potential method to determine the direction of the plane-of-the-sky magnetic field which is not available from the dust polarization or synchrotron emission.
That is, if we know the direction of either the line-of-sight electric current or the plane-of-the-sky magnetic field, we can determine the direction of the other.
In MHD models of outflows driven by the magneto-centrifugal mechanism, the electric current can flow either in the same direction as the outflow or in the opposite direction, depending on whether the magnetic field axis is anti-aligned or aligned to the rotation axis of the accretion disk\cite{1993WK}.
If the line-of-sight current of the 4A1 redshifted outflow is pointing away from us like the assumed pseudo-current, the positive slope in Fig.\ \ref{fig3} implies that the magnetic field counter-clockwisely (viewing from outflow pole to star) wraps around the outflow.
Alternately, if the line-of-sight current is pointing toward us, the slope in Fig.\ \ref{fig3} implies that the magnetic field clockwisely wraps around the outflow (see Methods for illustrations).
Although we cannot resolve the degeneracy at present, future Zeeman observations\cite{2024Mazzei} or rotation measurement analysis\cite{2025Rodriguez}, which probe the line-of-sight magnetic field direction, will resolve whether the magnetic field is wrapping clockwisely or counter-clockwisely.

Fig.\ \ref{fig4} shows the magnetic field structure inferred from the CO polarization.
The magnetic field segments are almost perpendicular to the outflow axis, indicating that the field is dominated by a toroidal component.
The $B_{pos}$ is stronger along the major axis of the outflow and becomes weaker at the edge of the outflow likely due to projection effect. 
At a distance about 300 au from 4A1, the magnetic field strength is greater than 3 mG and decreases to a strength less than 2 mG as the distance increases to 500 au. 
In the magneto-centrifugal launching mechanism of protostellar outflows, the magnetic field acts as the primary force to lift the gas with magnetic acceleration $a_B = \frac{J\times B}{c \mu m_H n_{H_2}}$ against the gravitational acceleration $a_G = \frac{-G M}{l^2}$ from the star, where $\mu = 2.86$ is the mean molecular weight\cite{2013Kirk}, $m_H$ is hydrogen atomic mass, $M$ is the mass of the star, and $l$ is the length of the outflow. %cite Shu and Pudritz?
Considering that $M \sim 1$ solar mass\cite{2008Goncalves} and the orders of magnitude for $B, n_{H_2}$, and $v{_d}$ are 1 mG, 10$^5$ cm$^{-3}$, and 10$^3$ cm s$^{-1}$ at $l \sim 400$ au, the $a_B \sim 10^{-4}$ cm s$^{-2}$ is about two orders of magnitude greater than the $a_G \sim 10^{-6}$ cm s$^{-2}$.
The dynamical time of accelerating outflow velocity from rest to reach 10 km s$^{-1}$ due to $a_B$ is approximately 200 years, consistent with the dynamical time independently estimated of gas traveling 400 au from rest accelerated by $a_B$.
Meanwhile, the magnetic pressure $\frac{B^2}{8 \pi} \sim 10^{-7}$ dyne cm$^{-2}$ is much higher than the thermal pressure $1.2 n_{H_2} k T \sim 10^{-9}$ dyne cm$^{-2}$ with a gas temperature $T \sim 70$ K of the outflow\cite{2012Yildiz}.
Thus, the toroidal magnetic fields are strong enough to accelerate and collimate the outflow.

According to the Kruskal-Shafranov criterion under ideal MHD conditions, an outflow becomes unstable to the helical kink mode when $\frac{B_p}{B_t} < \frac{l}{2 \pi r}$, where $B_p$ and $B_t$ are the poloidal and toroidal magnetic field strengths and $r$ is the radius of the outflow\cite{2008Moll}. For the IRAS 4A1 outflow, our wind-driven model fitting gives an estimated radius of $r \sim 80$ au at a length of $l \sim 400$ au. This implies that the magnetic field becomes unstable when the pitch angle of magnetic field, given by $\tan^{-1} \left( \frac{B_p}{B_t} \right)$, is less than $39^\circ$. The magnetic field segments in Fig.\ \ref{fig4} are almost perpendicular to the outflow axis, indicating a pitch angle well below this threshold and implying that the magnetic field structure in the IRAS 4A1 outflow is vulnerable to kink instability.
However, the growth of instability requires the Alfv\'en velocity $v_A = \frac{B_t}{\sqrt{4 \pi \mu m_H n_{H_2}}}$ being faster than the radial expansion velocity of the outflow\cite{2008Moll}. Considering that the orders of magnitude for $B$ and $n_{H_2}$ are 1 mG and 10$^5$ cm$^{-3}$, the corresponding $v_A$ of about 4.1 km s$^{-1}$ is slightly slower than the expansion velocity of about 4.4 km s$^{-1}$ estimated from our wind-driven {model}.

{The presence of dominant toroidal magnetic fields at a few hundred au scale in the IRAS 4A outflow, together with the dominant poloidal fields inferred from magneto-centrifugal wind modeling within the innermost $\sim$ 100 au of the IRAS 21078+5211 jet\cite{2022Moscadelli,2023OKb}, is consistent with the longitudinal evolution of magnetic fields predicted by magneto-centrifugal models where the magnetic field is predominantly poloidal inside the Alfv\'en surface and predominantly toroidal in the super-Alfv\'enic region. 
The super-Alfv\'enic state of the IRAS 4A outflow implied from the surpassing of the $\sim$ 10 km s$^{-1}$ outflow velocity over the Alfv\'en velocity $v_A \sim 4$ km s$^{-1}$ is also consistent with the kinematic condition for generating toroidal fields.
As our wind-driven model indicates that the IRAS 4A outflow is accelerated beyond several hundred au, the toroidal field may contribute to additional flow acceleration through $a_B$ at large distances.} 
In contrast to the HH 80–81 jet, which shows dominant poloidal magnetic fields at the edges of the jet at a scale of 0.4 pc {from a massive protostar\cite{2010Carrasco,2025Rodriguez}, our findings show that protostellar outflows from lower mass stars can have a predominantly toroidal magnetic configuration closer to the central protostar, indicating that they transition to the super-Aflv\'enic regime at smaller distances than their counterparts in more massive systems.}

Overall, the perpendicular alignment between the magnetic fields and outflow axis, the parallel alignment between the magnetic field structure and the rotational structure of the outflow, and the wind-driven parabolic features are consistent with the theoretical picture of toroidal magnetic fields in an outflow driven by the protostellar wind\cite{2006BP,2008Machida,2024Basu}. 
Similar to IRAS 4A, most of protostellar outflows are found to be misaligned to the magnetic fields of the cores\cite{2013Hull}, and our finding of toroidal magnetic fields supports the scenario of an outflow in a misaligned system driven by a magneto-centrifugal wind\cite{2015Seifried,2020Machida} rather than a spiral flow\cite{2017Matsumoto}. 
MHD simulations usually do not directly compute the electric current since in MHD, one deals directly with the magnetic field, and the curl of the magnetic field yields electric current. That is, electric current and $v_d$ are completely determined by the magnetic field distribution in the MHD simulations. 
If our interpretation of the correlation based on the Amp\`ere's law is correct, $v_d$ must be remarkably constant across the outflow and much slower than the drift velocity between ions and neutrals, providing a constraint for outflow simulations. 

\section{Methods}
\subsection{Observations and data reduction.}
Polarization observations of the NGC1333 IRAS 4A were carried out with ALMA in Band 6 at $\sim$ 230.5 GHz in Cycle 4, with 43 antennas in C40-5 configuration. 
The project number was 2016.1.01089.S. 
The phase center was ($\alpha_{2000},\delta_{2000}$) = (03$^h$29$^m$10.510$^s$, 31$^\circ$13'35.50'') toward the redshifted outflow of IRAS 4A.
%The field of view of linear polarization is Cycle 4 was 1/3 of the 25.3" primary beam. 
Three executions were carried out on 4 November 2016, with a total on-source time of 48.4 minutes.
The correlator was set up to have four spectral windows, with one for CO $J = 2-1$ at 230.538 GHz, one for C$^{18}$O $J = 2-1$ at 219.560 GHz, one for SiO $J = 5-4$ at 217.105 GHz, and one for the continuum at 234 GHz. Our dust polarization results have been published before\cite{2020Ko}.
In this paper, we present CO polarization results. 
The velocity resolution of the CO spectral window is 0.159 km s$^{-1}$. To increase the sensitivity for CO polarization measurements, we binned 20 channels to produce the images presented in this work at a velocity resolution of 3.175 km s$^{-1}$.

The data were calibrated using the CASA package (versions 4.7.2) with Quasar J0238+1636 as a flux density scale calibrator, Quasar J0238+1636 as a passband calibrator, Quasar J0336+3218 as a complex gain calibrator, and Quasar J0334-4008 as a polarization calibrator.
The continuum emission in the channels with CO emission was subtracted using line-free channels.
A phase-only self-calibration of the data was performed to improve the map fidelity using the CO Stokes $I$ emission.
We used natural weighting to generate the CO polarization maps and dust emission (including Stokes $I$, $Q$, and $U$ parameters) at a resolution of $0.49” \times 0.33”$ with a position angle of $-7.9^\circ$.
The CO Stokes $I$ maps (the intensity maps in Fig.\ \ref{fig1}, the moment 0 map, moment 1 map, and PV diagrams in Fig.\ \ref{fig2}, and the number density maps of Fig.\ \ref{fig4}) were generated with a pixel size of 0.05".  
In order to derive the curl of magnetic field from polarization segments at sufficient angular displacement, the CO polarization measurements (the black segments in Figs.\ \ref{fig1} and \ref{fig2}, and the magnetic field segments in Fig.\ \ref{fig4}) were generated using MIRIAD to bin the Stokes $I$, $Q$, and $U$ maps from 0.05" pixel size to 0.4" pixel size (pixel area $\sim$ 0.9 beam area).
The noise level is 4.8 mJy beam$^{-1}$ in the Stokes $I$ maps and 0.92 mJy beam$^{-1}$ in the Stokes $Q$ and $U$ maps for the CO line.
All the polarization segments in this work satisfy the criteria of Stokes $I$ emission higher than 5 signal-to-noise ratio and polarized emission  higher than 3 signal-to-noise ratio.

Extended Data Figure 1 shows the CO $J = 2-1$ polarization channel maps with length of polarization segments proportional to the polarization percentage.
No primary beam correction is applied to the maps because according to ALMA Technical Handbook in Cycle 4\cite{ALMA4}, we only focus on the region within 1/3 of the primary beam where the polarized emission can be mapped properly.
The maximum recoverable scale of our data is 2.8". Considering that the lowest few contours of CO Stokes $I$ emission have structures larger than the maximum recoverable scale, the Stokes $I$ emission at those contours is partially resolved out. Because the polarized emission is less extended than the Stokes $I$ emission, the polarized emission should be less resolved out than the Stokes $I$ emission, producing a few polarization segments with polarization percentages larger than 10\% between the lowest and the second lowest contours of the outflow. The analysis in this work only utilizes the polarization angle, not the polarization percentage, and hence our results would not be affected by this missing flux issue. 

In this work, we use the term `orientation' to represent the polarization angle, which ranges from -90$^\circ$ to +90$^\circ$, and the term `direction' to represent the magnetic field direction, which ranges from -180$^\circ$ to +180$^\circ$. For the polarization measurement with orientation in the figures, we call it `segment'. For the magnetic field measurement with direction in the figures, we call it `vector'.

%minimum p% is 0.5% 
\subsection{Fitting of wind-driven model.}
In the wind-driven model, molecular outflows are the ambient material swept-up by the wide-angle wind from a young star, and the intersection of the wind and the ambient gas determines the shape and velocity of the outflow.
For a constant-velocity protostellar wind with its mass-loss rate proportional to $\sin^{-2}\theta$ blowing into an ambient medium with density profile proportional to $\rho^{-2} \sin^{2}\theta$, the outflow has an approximately parabolic shape in which the velocity increases with the distance from the star, where $\rho$ and $\theta$ are the radial distance and polar angle in spherical coordinate system\cite{1996LS}.
As a result, the shell of the outflow cavity produces parabolic structure in PV diagram.

In the cylindrical coordinate system, the structure and velocity of the outflow can be simplified with an empirical equation\cite{2000Lee}:
\begin{equation}
Z = CR^2,\,	v_R = v_0 R,\,	v_Z = v_0 Z,
\end{equation}
where $Z$ is the distance along the outflow axis; $R$ is the radial size of the outflow perpendicular to the outflow axis; $C$ and $v_0$ are free parameters that describe the spatial and velocity distributions of the outflow shell. 
We use the emission peaks in the PV diagram to fit the model (Extended Data Figure 2).
The parabolic features of the 4A1 redshifted and blueshifted high-velocity outflows are reproduced by a model with small opening angle of $C_{high} = 17 \pm 13$ arcsec$^{-1}$, $v_0$ = $15.6 \pm 7.5$ km s$^{-1}$ arcsec$^{-1}$, and an inclination angle $\theta_{incl} = 18.1^\circ \pm 7.8^\circ$ with respect to the plane of the sky.
Using the same $v_0$ and inclination angle, the 4A1 redshifted low-velocity outflow is reproduced with a large opening angle of $C_{low} = 1.7 \pm 1.5$ arcsec$^{-1}$.
The solid and dashed curves in Fig.\ \ref{fig2}d and Extended Data Figure 2 show the fittings of small and large opening angles, respectively.

\subsection{Estimation of magnetic field strength.}

We adopt the Davis--Chandrasekhar--Fermi equation\cite{1951Davis, 1953CF} to derive the plane-of-the-sky magnetic field strength {and its $x$ and $y$ components:}
\begin{equation}
B_{pos} = F \sqrt{4 \pi \mu m_H n_{H_2}}(\delta v_{los}/\delta \phi),
%\end{equation}
%\begin{equation}
\quad
%\color{red}
B_x = B_{pos} \sin \theta_{PA},
\quad
B_y = B_{pos} \cos \theta_{PA},
\end{equation}
where $\mu = 2.86$ is the mean molecular weight in nearby molecular clouds assuming that the gas is 70\% H$_2$ by mass\cite{2013Kirk}; $m_H$ is the atomic mass of hydrogen; $n_{H_2}$ is the gas number density; $\delta v_{los}$ is the line-of-sight velocity dispersion, and we use a value of 0.8 km s$^{-1}$ from the C$^{18}$O turbulent velocity of IRAS 4A outflows\cite{2012Yildiz}; $\delta\phi$ is the dispersion of CO polarization angles by counting adjacent polarization segments (e.g.\ the ``Unsharp Masking'' method\cite{2017Pattle}); $F$ is an order unity factor, which is typically taken as $F \approx 0.5$ (ref.\ \cite{2001Ostriker}); {and $\theta_{PA}$ is the CO polarization angle measured on the plane of the sky from north toward east in the convention of International Astronomical Union.}

To derive $n_{H_2}$, we adopt Eqs.\ 34, 52, 66, and 82 in Ref.\ 33 of calculating CO column density ($N_{CO}$) under optically thin condition with Rayleigh-Jeans approximation:
\begin{equation}
N_{CO} = \left(\frac{3k}{8 \pi^3 \nu J_u \mu_{CO}^2}\right)\left(\frac{kT_{ex}}{hB_{CO}}+\frac{1}{3}\right)\rm exp\left(\frac{E_u}{T_{ex}}\right)\int T_{CO} dv,
\end{equation}
where $\nu = 230.538$ GHz is the rest frequency of CO $J = 2-1$ transition; $J_u = 2$ is the total angular momentum quantum number of the upper level; $\mu_{CO} = 0.11$ Debye is the dipole moment of CO molecule; $T_{ex}$ is the excitation temperature, and we use the value of 70 K from the observed CO rotational temperature of $69 \pm 7$ K in the IRAS 4A outflows\cite{2012Yildiz}; $B_{CO} = 57635.968$ MHz is the rigid rotational constant of CO molecule; $E_u = 16.59608$ K is the upper level energy of CO $J = 2-1$ transition; and $T_{CO}$ is the brightness temperature of CO Stokes $I$ emission. Next, H$_2$ column density ($N_{H_2}$) is derived with $N_{H_2} = N_{CO}/[CO/H_2]$ using the canonical value of CO abundance $[CO/H_2] \approx 10^{-4}$ in molecular clouds\cite{1978Dickman}. Considering that the line-of-sigh depth of a velocity channel $L = \Delta v/(v_0 \sin \theta_{incl}) d$ using channel velocity width $\Delta v = 3.175$ km s$^{-1}$, $v_0$ and $\theta_{incl}$ from the fitting of a wind-driven model, and IRAS 4A distance $d = 300$ pc (the distance of the NGC 1333 cloud from the Earth is 293 $\pm$ 22 pc calculated from trigonometric parallaxes\cite{2018Ortiz} or 299 $\pm$ 14 pc calculated later from a combination of stellar photometry, astrometric data, and CO spectral-line maps\cite{2018Zucker}
and we adopt the value of 300 pc in this work) 
, we obtain $n_{H_2} = N_{H_2}/L$ {with $L \sim 198$ au which is comparable to the width of CO emission  in Fig.\ \ref{fig1}. Note that $v_0$ is in units of km s$^{-1}$ arcsec$^{-1}$, and the dimension of arcsec converts $d$ in pc to $L$ in au when calculating $L = \Delta v/(v_0 \sin \theta_{incl}) d$}. Extended Data Figures 3, 4, 5, and 6 show the maps of $n_{H_2}$, $\delta \phi$, $B_{pos}$, and $\nabla \times B_{pos}$ obtained from the channels maps in Fig.\ \ref{fig1}. 

{
Applications of the Davis--Chandrasekhar--Fermi method to the GK effect have been reported for SiO emission in protostellar outflows and jets\cite{2018Lee,2020Hirota}.
In addition, the structure function approach of the Davis--Chandrasekhar--Fermi method with the GK effect in CO lines has been applied to outflowing gas around asymptotic giant branch stars\cite{2023VT,2024Vlemmings}.
Recently, a modified Davis--Chandrasekhar--Fermi method\cite{2021LR} has been proposed to account for the presence of a large-scale velocity field $U_0$ that is parallel to the local magnetic field direction with} 
\begin{equation}
{B_{pos,mDCF} = B_{pos} \big | 1 - \delta \phi \frac{U_0}{\delta v_{los}} \big |.}
\end{equation}
{This modified method has been used to estimate magnetic field strengths from dust polarization and GK effect in galactic outflows and circumnuclear disks\cite{2021LR,2023Guerra,2026LR}. 
In most cases, $B_{pos,mDCF}$ is weaker than $B_{pos}$ because the  correction factor $| 1 - \delta \phi \frac{U_0}{\delta v_{los}} \big |$ subtracts the magnetic field structure generated by ordered flows from the polarization angle dispersion.
Under typical flow conditions of $\delta \phi < 1$ and $U_0 \sim \delta v_{los}$, this correction factor is less than unity.}

%Pattle PPVII paper that DCF methods is systematically overestimated.
%DCF on GK effect observation
%Vlemmings 2024: outflow of AGB stars CO; Structure function analysis
%Vlemmings & Tafoya 2023: outflow of AGB stars CO; Structure function analysis
%Hirota 2020: Orion SiO outflow; DCF 
%Chin Fei : SiO outflow; DCF
%Cortes: CS envelope gas optically thick; DCF 
%Enrique Lopez-Rodriguez 2025: CO outflow AGN; DCF and Modified DCF

{To evaluate how the modified Davis--Chandrasekhar--Fermi method affects our results, we consider the rotational structure of the outflow as the source of $U_0$, given that the magnetic field inferred from CO polarization is parallel to the rotational structure of the outflow.
At the first knot of the redshifted outflow, the velocity gradient is $4.9 \times 10^{-2}$ km s$^{-1}$ au$^{-1}$, corresponding to a rotational velocity of 4.0 km s$^{-1}$ for the high-velocity shell at a projected distance $Z = 400$ au. We further assume that the rotational velocity of the shell scales linearly with $Z$, and we adopt the plane-of-the-sky component of the rotational velocity as $U_0$.
Extended Data Figures 7, 8, 9, and 10 show the maps of $U_0$, $v_A$, $B_{pos,mDCF}$, and $\nabla \times B_{pos,mDCF}$ of the four velocity channels. Overall, because the regions exhibiting CO polarization satisfy the conditions $\delta \phi < 1$ and $U_0 \sim \delta v_{los}$, $B_{pos,mDCF}$ is slightly weaker than $B_{pos}$.
Extended Data Figures 11 presents the same analysis as Fig.\ \ref{fig3} with $B_{pos,mDCF}$ substituted for $B_{pos}$.
A linear fitting of Eq.\ 2 gives a slope $v_d = (5.8 \pm 1.5) \times 10^{2}$ cm s$^{-1}$ and an intercept $\frac{1}{c} \frac{\partial E_z}{\partial t} = (-7.6 \pm 2.5) \times 10^{-19}$ cm$^{-3/2} $ g$^{1/2}$ s$^{-1}$.
The slope derived using $B_{pos,mDCF}$ is consistent with the slope derived using $B_{pos}$, indicating that our estimation of $v_d$ is robust. Because the $B_{pos,mDCF}$ is generally weaker than $B_{pos}$, the intercept is smaller compared to the intercept in Fig.\ \ref{fig3}.}

%\subsubsection{Estimation of Xe, vd}

\subsection{Correlation assuming $v_d$ proportional to outflow velocity.}
In addition to considering a constant $v_d$ in the main text, we also consider that $v_d$ is proportional to outflow velocity. In this case, $J_z = {\varrho v_d = \varrho v_{CO} X_d}$, where $v_{CO}$ is the velocity of CO channel with respect to the IRAS 4A systemic velocity, and $X_d$ is a unit-less factor describing the radio between $v_d$ and $v_{CO}$.
Hence, Eq.\ 1 can be rewritten as
%\frac{\partial B_y}{\partial x}-\frac{\partial B_x}{\partial y} = (\frac{4 \pi e}{c} \sqrt{2} \times 10^{-5} n_{H_2}^{1/2} v_{CO}) X_d + \frac{1}{c} \frac{\partial E_z}{\partial t}.
\begin{equation}
\frac{\partial B_y}{\partial x}-\frac{\partial B_x}{\partial y} = (\frac{4 \pi}{c} {\varrho} v_{CO}) X_d + \frac{1}{c} \frac{\partial E_z}{\partial t}.
\end{equation}

Extended Data Figure 12 shows the correlation between the $\frac{\partial B_y}{\partial x}-\frac{\partial B_x}{\partial y}$ derived from CO linear polarization and the $\frac{4 \pi}{c} {\varrho} v_{CO}$ derived from CO total emission. 
A linear fitting of Eq.\ 6 gives a {slope} $X_d = (3.0 \pm 1.7) \times 10^{-4}$ and an intercept $\frac{1}{c} \frac{\partial E_z}{\partial t} = (-2.5 \pm 2.0) \times 10^{-19}$ cm$^{-3/2} $ g$^{1/2}$ s$^{-1}$.
For the 4A1 outflow velocity of 10 km s$^{-1}$, the $v_d = v_{CO} X_d$ is about $3 \times 10^2$ cm s$^{-1}$, in agreement with the result when assuming a constant $v_d$. 
However, the fitting goodness of $X_d$ is less than 1.8$\sigma$, less significant than the 3.3$\sigma$ goodness of $v_d$ when assuming a constant $v_d$, indicating that $v_d$ in outflow is more likely to be a constant than proportional to outflow velocity.

\subsection{Fitting robustness of $v_d$ and $\frac{1}{c} \frac{\partial E_z}{\partial t}$.}
{Despite its low statistical significance, the intercept in Fig.\ \ref{fig3} of $\frac{1}{c} \frac{\partial E_z}{\partial t} = (-8.5 \pm 2.9) \times 10^{-19}$ cm$^{-3/2}$ g$^{1/2}$ s$^{-1}$ exceeds the displacement current expected in ideal MHD (estimated from $E = -\frac{v \times B}{c}$ with $v \sim$ 10 km s$^{-1}$ and $B \sim$ 1 mG in a dynamical time of 200 years) by at least eight orders of magnitude.
We speculate that this discrepancy mainly comes from two systematic effects in the derivation of the data points in Fig.\ \ref{fig3}.
First, the intercepts in Fig. \ref{fig3}, Extended Data Figure 11, and Extended Data Figure 12 vary by a factor of 3, indicating a strong dependence of the intercept on the  assumptions involved in the Davis--Chandrasekhar--Fermi method and the source of $J_z$. 
Second, the unsharp masking method used to calculate $\delta \phi$ can systematically bias the $\frac{\partial B_y}{\partial x}$ and $\frac{\partial B_x}{\partial y}$ data toward negative values in Fig.\ \ref{fig3}.} 

{Extended Data Figure 13 presents a simulation illustrating how the unsharp masking method can generate a systematic offset in the intercept.
Our simulation takes four steps: 
(1) we generate input magnetic fields identical to the test models used in the original work introducing the unsharp masking method (see Appendix B in ref.\ \cite{2017Pattle}) with parabolic curvatures in the form $y = \frac{1}{4f}x^2$ for a range of focal lengths $f$.
With the magnetic field axis aligned along the $y$ axis, the case of $f = \infty$ represents pure edge-on toroidal fields, while smaller values of $f$ mimic magnetic fields with stronger poloidal components.
Extended Data Figures 13a shows an example model with $f = 1.5$. 
The tests in ref.\  \cite{2017Pattle} determined polarization orientations from input models but did not specify magnetic field strengths. 
Here we assume that the input magnetic field strengths follow the Davis--Chandrasekhar--Fermi method with $B_{pos}$  proportional to  $1/\delta \phi$ and adopt a spatial distribution in which the field is strongest near the center $(x = 0, y = 0)$ and decreases with increasing distance, similar to the distribution of observed magnetic field strengths.
(2) We calculate the angular dispersion $\delta \phi_{mod}$ using the variation of polarization orientations inferred from input magnetic fields in a box of $3 \times 3$ pixels (Extended Data Figures 13b).
(3) In addition to the initial $\delta \phi$, the angular dispersion $\delta \phi_{mod}$ arising from unsharp masking method leads to a modified field strength $B_{pos,mod} = B_{pos}\frac{\delta \phi}{\sqrt{\delta \phi^2 + \delta \phi_{mod}^2}}$. Adopting $\delta \phi = 1/6$ radians which is comparable to the typical polarization angle uncertainty for polarization measurements with a signal-to-noise ratio of 3 (ref.\ \cite{1974WK}), Extended Data Figures 13c shows the output $B_{pos,mod}$. The $B_{pos,mod}$ vectors are systematically shorter (i.e.\ weaker in strength) than the $B_{pos}$ vectors, and this modification also alters the field gradients.  
(4) We compare the field gradients derived from $B_{pos,mod}$ to those derived from $B_{pos}$ with $f$ = 1.2, 1.5, and 3.0, spanning the curvature range of the observed magnetic field structure.
Extended Data Figures 13d shows that the $\frac{\partial B_y,mod}{\partial x} - \frac{\partial B_x,mod}{\partial y}$ data points are systematically smaller than the $\frac{\partial B_y}{\partial x} - \frac{\partial B_x}{\partial y}$ data points of the input models. For $f = \infty$, the input magnetic fields are perfectly horizontal, producing zero $\delta \phi_{mod}$ and hence identical output field gradients as those of the input data. 
As $f$ decreases, the deviation between the output and input gradients increases.
For $f$ = 1.2 and 1.5, the deviation becomes comparable to the magnitude of the data, similar to the ratio between the intercept and the magnitude of $y$-axis data in Fig.\ \ref{fig3}. 
We therefore attribute the large interception in Fig.\ \ref{fig3} to systematic biases rather than a physical displacement current in protostellar outflow.}

%In this simulation, the initial magnetic fields follow the Davis--Chandrasekhar--Fermi method that $B_{pos}$ is proportional to  $1/\delta \phi$. 
%Because the unsharp masking method assumes locally uniform magnetic field, spatial variation of magnetic field structure introduces an additional angle dispersion $\delta \phi_{mod}$, leading to a modified field strength $B_{pos,mod} = B_{pos}\frac{\delta \phi}{\sqrt{\delta \phi^2 + \delta \phi_{mod}^2}}$.
%Assuming $\delta \phi = 1/6$ radians as the typical error in polarization angle for polarization measurements with a signal-to-noise ratio of 3 (ref.\ \cite{1974WK}), the resulting $B_{pos,mod}$ vectors are systematically weaker than the $B_{pos}$ vectors (Extended Data Figures 13a and 13c). 
%This modification also alters the field gradients: for example, the $B_x$ in Extended Data Figures 13a is identical at $x = 0$ and hence $\frac{\partial B_x}{\partial y} = 0$, but the modified $B_{x,mod}$ at $x = 0$ in Extended Data Figures 13c considerably increases with $y$, generating a positive $\frac{\partial B_x,mod}{\partial y}$ and shifting the data point $\frac{\partial B_y,mod}{\partial x} - \frac{\partial B_x,mod}{\partial y}$ toward a negative value in Extended Data Figures 13d. 
%Similar shifts occur for all simulated data points. We therefore attribute the large interception in Fig.\ \ref{fig3} to systematic biases rather than a displacement current in a protostellar outflow.%}

{In contrast, the fitting of $v_d$ is more robust than the intercept. The values of $v_d$ in Fig.\ \ref{fig3} and Extended Data Figure 11 are highly consistent with fitting goodness both higher than 3$\sigma$. Moreover, the slopes in Extended Data Figure 13d for models with $f$ = 1.2, 1.5, and 3 are comparable to that of the input models, indicating that the unsharp masking method does not modify $v_d$ significantly. We therefore interpret $v_d$ as a meaningful physical quantity.}

Note that according to Eq.\ 4, the adopted values of $F$, $\mu$, and $\delta v_{los}$ can systematically scale the derived $B_{pos}$, i.e.\ systematically scaling up or down the data points in Fig.\ \ref{fig3}. 
Consequently, the fitted slope $v_d$ and intercept $\frac{1}{c} \frac{\partial E_z}{\partial t}$ from Fig.\ \ref{fig3} are also scaled by these adopted parameters, typically within a factor of two (primarily influenced by $F$).
However, the fitting goodness of the slope and intercept would not be changed by those adopted values since the error bars of the data points in Fig.\ \ref{fig3} are scaled proportionally with $B_{pos}$.
Additionally, since both sides of Eq.\ 2 contain a factor of $n_{H_2}^{1/2}$ ($B_x$ and $B_y$ are implicitly proportional to $n_{H_2}^{1/2}$), our adopted values of $T_{ex}$, $[CO/H_2]$, $v_0$, $\theta_{incl}$, and $d$ used in deriving $n_{H_2}$ have weak impact on the values and fitting goodness of the slope and intercept. %Therefore, our adopted values of physical and chemical parameters when deriving $B_{pos}$ and $n_{H_2}$ have minimal influence on both the magnitude and the fitting robustness of the results in Fig.\ \ref{fig3}.

\subsection{Illustration of the $B_{pos}$ direction inferred from the analysis of the Amp\`ere's law.}
If the line-of-sight current of the 4A1 redshifted outflow is pointing away from us, the positive slope in Fig.\ \ref{fig3} indicates that magnetic fields counter-clockwisely wrap around the outflow as viewing from outflow pole to star. Extended Data Figure 14 illustrates the inferred vectors of counter-clockwise toroidal magnetic fields.
The magnetic field vectors in the velocity channel at $+3.090$ km s$^{-1}$ are in the southwestern direction because along the line of sight, the low-velocity channel traces the foreground layer of the outflow cavity (the blue box in Extended Data Figure 2). On the contrary, the high-velocity channels from $+6.265$ to $+12.615$ km s$^{-1}$ reveal that the magnetic field vectors in the background layer (the red box in Extended Data Figure 2) are in the northeastern direction. 

If the line-of-sight current of the 4A1 redshifted outflow is pointing toward us, the x-axis values of the data points in Fig.\ \ref{fig3} need to be multiplied by $-1$, resulting in a negative slope. Since the right-hand rule requires a positive slope, the magnetic field vectors shown in Extended Data Figures 3, 4, 5, and 6 must be rotated by 180$^\circ$. This adjustment implies that the magnetic fields clockwisely wrap around the outflow as viewing from outflow pole to star. Extended Data Figure 15 illustrates the inferred vectors of clockwise toroidal magnetic field.
Extended Data Figure 16 shows the 3-dimensional geometry of the outflow and magnetic fields for the two scenarios of the line-of-sight current pointing away from or toward us.

\subsection{Data availability}
ALMA data are available at https://almascience.nrao.edu/aq/.
The polarization channel maps in Fig.\ \ref{fig1} are available at https://github.com/taochung/COpol.

\subsection{Code availability}
The codes analyzing the polarization channel maps are available at https://
github.com/taochung/COpol.

\clearpage

\begin{figure}[h!]
\centering
%\hspace{-0.5cm}
\includegraphics[width=0.9\textwidth]{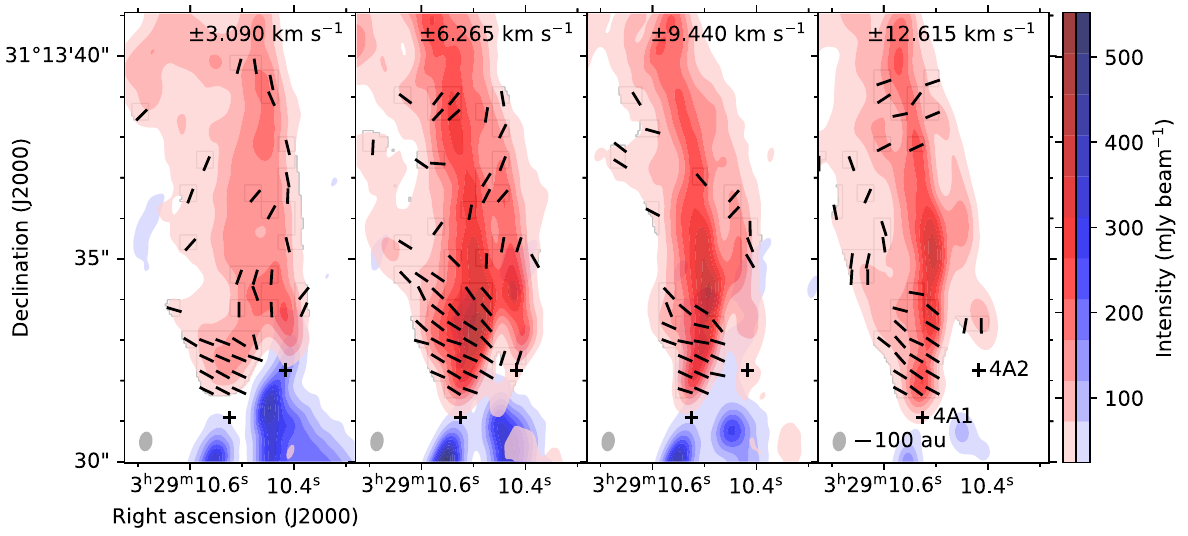}
\caption{ {\bf IRAS 4A CO $J = 2-1$ polarization maps at velocity channels of 3.175 km s$^{-1}$ intervals.} 
Red and blue filled contours represent the total emission of the redshifted and blueshifted outflows, respectively.
Contours start from 5$\sigma_I$ in steps of 10$\sigma_I$ with $\sigma_I$ = 4.8 mJy beam$^{-1}$, which is the noise level of total emission (Stokes I).
The central channel velocities with respect to the systemic velocity ($v_{LSR}$ = 6.86 km s$^{-1}$) of 4A1\cite{2019Su} and the synthesized beam are indicated in each panel.
Black segments show the polarization orientations of the redshifted emission with an identical length. 
A length scale of 100 au and crosses for the positions of 4A1 and 4A2 are marked in the highest velocity channel.}
\label{fig1}
\end{figure}
\clearpage

\begin{figure}[h!]
\centering
\includegraphics[width=1.0\textwidth]{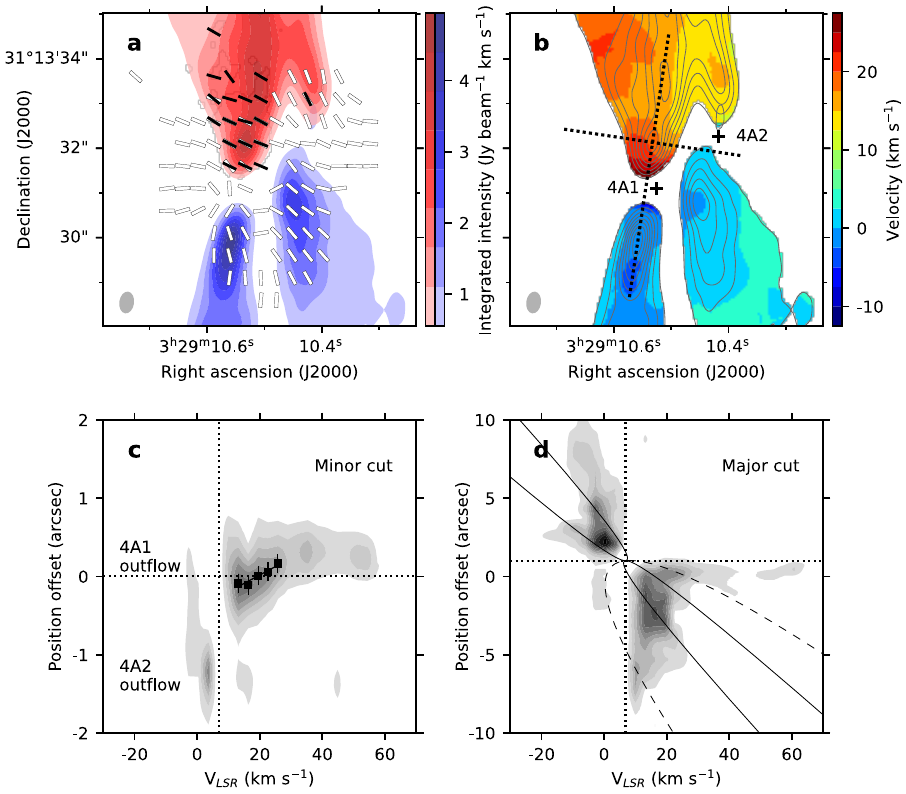}
%\label{fig2}
\end{figure}

%%%%%%%%%
%QZ: also add labels '4A1' and '4A2' next to '+'
\begin{center}
\setbox0\vbox{\makeatletter
\let\caption@rule\relax
\captionof{figure}{{\bf Velocity-integrated intensity map (moment 0), intensity-weighted mean velocity map (moment 1), and PV diagrams of the IRAS 4A CO outflows.} 
{\bf a}, Velocity-integrated polarization map with filled contours from 5$\sigma_{int}$ in steps of 5$\sigma_{int}$ with $\sigma_{int}$ = 0.11 Jy beam$^{-1}$ km s$^{-1}$. Black and white segments show the polarization orientations of the CO redshifted emission and the magnetic field orientations inferred from the dust polarization, respectively. The length of the segments is unified.
{\bf b}, Color image of the velocity field overlapped with the contours of the CO integrated emission. The dotted lines denote the cuts of the PV diagrams passing the major axis of the 4A1 outflow and the minor axis of the first knot in the 4A1 redshifted outflow.   
{\bf c}, PV diagram along the minor cut with filled contours at the same levels as those in Fig.\ \ref{fig1}. The horizontal and vertical dotted lines indicate the peak position of the first knot and the systemic velocity of 4A1.
The squares mark the positions of emission peaks above the fourth contour with error bars showing the uncertainties in the peak positions, as determined from Gaussian fits. The solid line marks the linear velocity gradient across the first knot.
{\bf d}, PV diagram along the major cut with filled contours at the same levels as those in Fig.\ \ref{fig1}. The horizontal and vertical dotted lines indicate the projected position of 4A1 on the PV cut and the systemic velocity of 4A1. The solid and dashed curves denote the wind-driven models of the 4A1 outflow.}
\label{fig2}
\global\skip1\lastskip\unskip
\global\setbox1\lastbox
}
\unvbox0
\setbox0\hbox{\unhbox1\unskip\unskip\unpenalty
\global\setbox1\lastbox}
\unvbox1
\vskip\skip1
\end{center}
%%%%%%%%%

\begin{figure}[h!]
\centering
\includegraphics[width=0.9\textwidth]{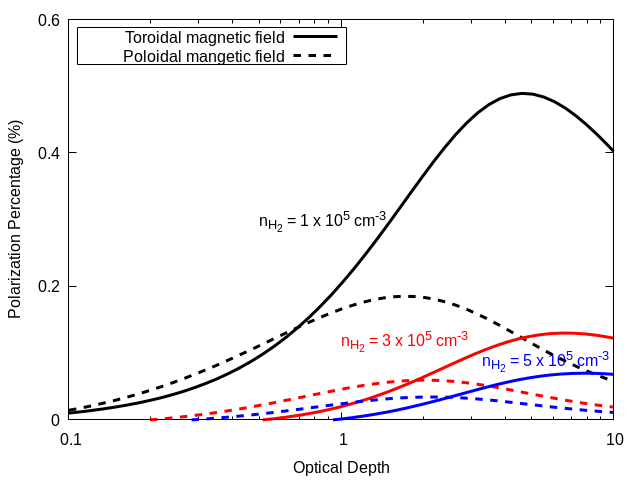}
\caption{{{\bf Linear polarization percentage of the CO $J = 2–1$ GK effect as a function of optical depth.} We consider gas temperature of 70 K, background temperature of 2.7 K, and inclination angle of 18$^\circ$ for the GK effect model. The solid and dashed lines represent the GK effect arising from toroidal and poloidal magnetic fields, respectively.
The black lines, red lines, and blue lines represent the GK effect at $n_{H_2} = 1 \times 10^5$, $3 \times 10^5$, and $5 \times 10^5$ cm$^{-3}$, respectively.}}
\label{fig3new}
\end{figure}

\begin{figure}[h!]
\centering
\includegraphics[width=0.9\textwidth]{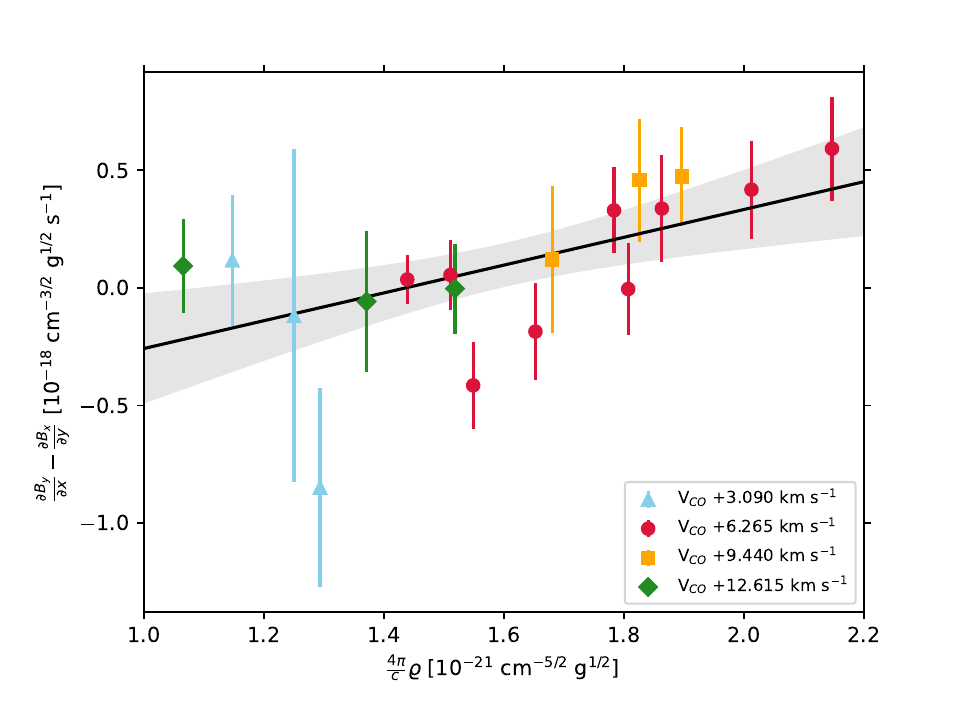}
\caption{{\bf The correlation between curl of magnetic field strength and pseudo-electric current {density}.} The pseudo-electric current {density} inferred from CO total emission is plotted in the $x$ axis, and the curl of $B_{pos}$ inferred from CO linear polarization is plotted in the $y$ axis. The cyan triangles, red dots, orange squares, and green diamonds show the data points with error bars (1$\sigma$) derived from the four velocity channels in Fig.\ \ref{fig1}. 
The black line shows the linear regression of Eq.\ 2 with gray region marking the 95\% confidence interval of the linear regression.} %{\bf a}, The correlation considering parallel orientation between CO polarization and magnetic field. {\bf b}, The correlation considering perpendicular orientation between CO polarization and magnetic field.}
\label{fig3}
\end{figure}

\begin{figure}[h!]
\centering
\includegraphics[width=0.9\textwidth]{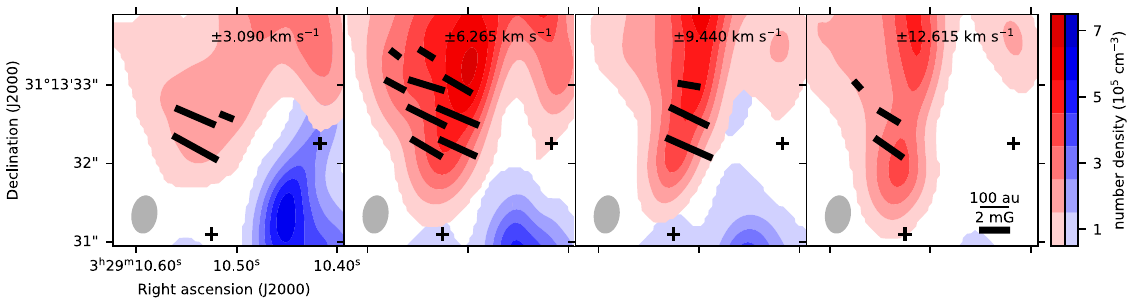}
\caption{{\bf Channel maps of magnetic field segments in the 4A1 redshifted outflow.} 
Red and blue filled contours represent the gas number density of the redshifted and blueshifted outflows with levels starting from 5 $\times$ 10$^4$ cm$^{-3}$ in steps of 10$^5$ cm$^{-3}$. The black segments show the magnetic field orientation with length proportional to the field strength.
The labels of channel velocities, synthesized beam, length scale, and crosses are the same as those in Fig.\ \ref{fig1}.}
\label{fig4}
\end{figure}

\clearpage

\bibliographystyle{naturemag}
\bibliography{references1_NA}

\begin{addendum}
\item 
This paper makes use of the following ALMA data: ADS/JAO.ALMA$\#$2016.1.01089.S.
ALMA is a partnership of ESO (representing its member states), NSF (USA) and NINS (Japan), together with NRC
(Canada), MOST and ASIAA (Taiwan), and KASI (Republic of Korea), in cooperation with the Republic
of Chile. The Joint ALMA Observatory is operated by ESO, AUI/NRAO and NAOJ.
The National Radio Astronomy Observatory is a facility of the National Science Foundation operated under cooperative agreement by Associated Universities, Inc.
J.M.\ G.\ acknowledges support by the grant PID2023-146675NB-I00 (MCI-AEI-FEDER, UE).
{This work is also partially supported by the program Unidad de Excelencia María de Maeztu CEX2020-001058-M.}
D.\ L.\ is a New Cornerstone Investigator.
D.\ L.\ was supported by National Nature Science Foundation of China (NSFC) under Grant No.\ 11988101, 12003047 and 12173053.

\item[Author Contributions] 
T.-C. C.\ initiated the ALMA project with S.-P.\ L., Q.\ Z., J.M.\ G., R.\ R., Z.-Y.\ L., and C.-F.\ L.
T.-C.\ C.\ reduced the data in consultation with E.\ M.\ and J.M.\ G.
All authors analyzed and discussed the observations and contributed to the paper.
  
\item[Competing Interests] 
The authors declare that they have no competing financial interests.

\end{addendum}
\clearpage

\renewcommand{\figurename}{Extended Data}

\renewcommand{\thefigure}{Figure 1}
\begin{figure*}
\centering
\includegraphics[width=1.\textwidth]{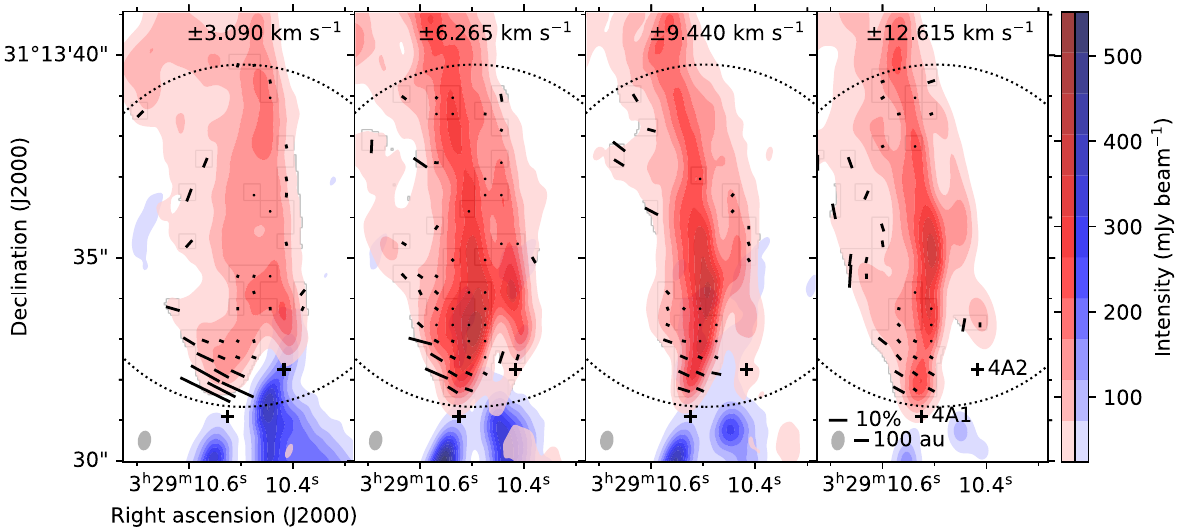}
\caption{IRAS 4A CO $J = 2-1$ polarization maps with length of polarization segments proportional to the polarization percentage. A scale bar of 10\%  polarization percentage is shown in the highest velocity channel. Dotted circles show the field of view of linear polarization which is 1/3 of the 25.3" primary beam. The contours, labels of channel velocities, synthesized beam, length scale, and crosses are the same as those in Fig.\ \ref{fig1}.}
\end{figure*}

\renewcommand{\thefigure}{Figure 2}
\begin{figure*}
\centering
\includegraphics[width=1.\textwidth]{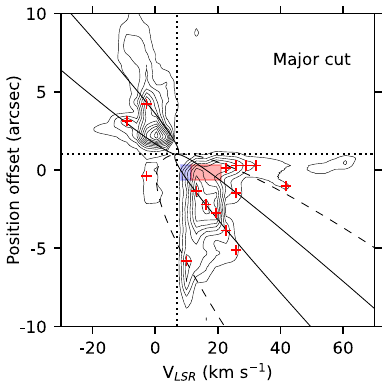}
\caption{PV diagram along the major cut with the fitting of wind-driven model. The contours, dotted lines, solid curve, and dashed curve are the same as those in Fig.\ \ref{fig2}d. Red crosses mark the emission peaks used for the fitting. The blue box marks the region of magnetic field segments in the velocity channel at 3.090 km s$^{-1}$ in Fig.\ \ref{fig4}, which probes the foreground layer of the outflow. The red box marks the regions of magnetic field segments in the velocity channels from 6.265 km s$^{-1}$ to 12.615 km s$^{-1}$ in Fig.\ \ref{fig4}, which probes the background layer.}
\end{figure*}

\renewcommand{\thefigure}{Figure 3}
\begin{figure*}
\centering
\includegraphics[width=1.\textwidth]{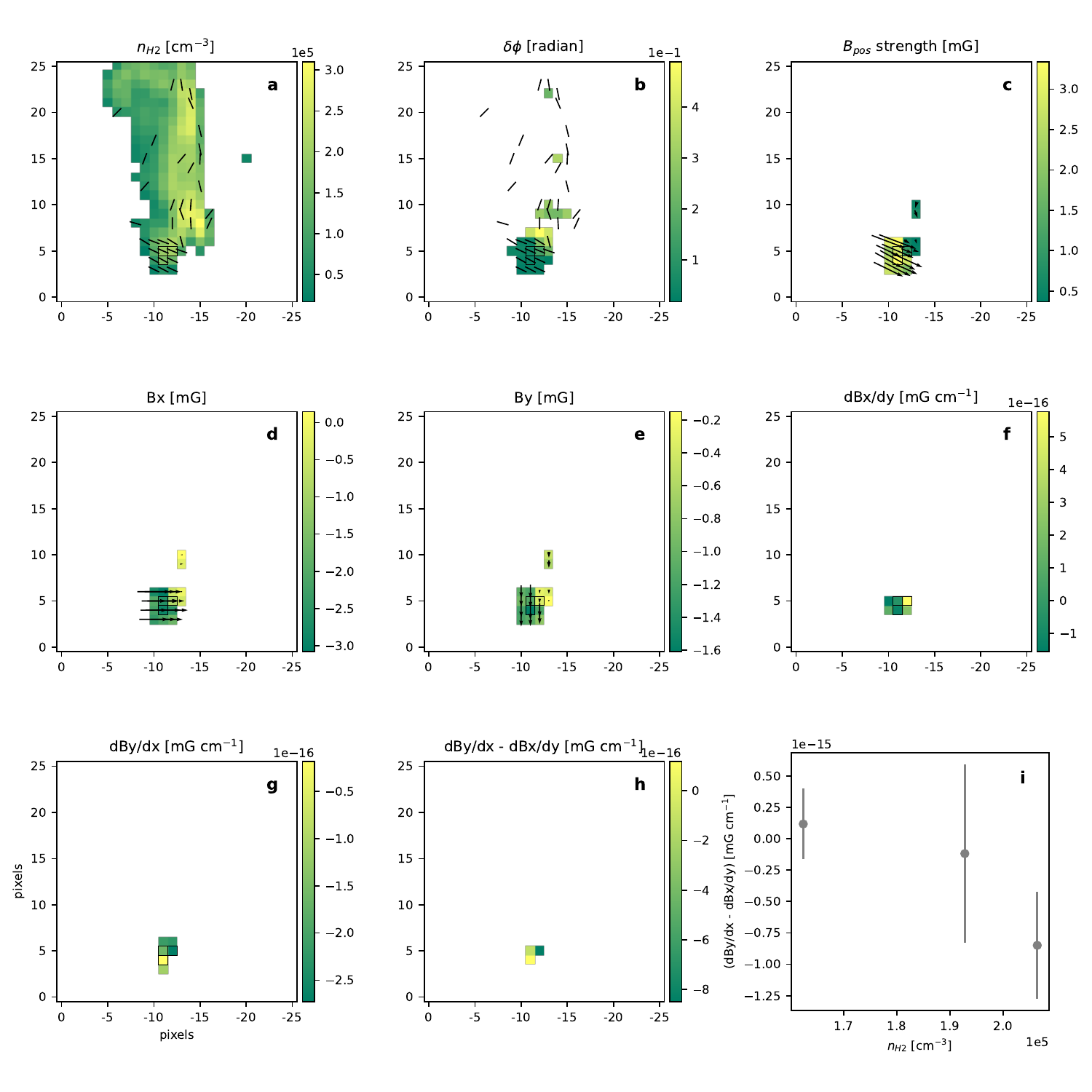}
\caption{Maps of (a) $n_{H_2}$, (b) $\delta \phi$, (c) $B_{pos}$, (d) $B_x$, (e) $B_y$, (f) $\partial B_x$/$\partial y$, (g) $\partial B_y$/$\partial x$, (h) $\nabla \times B_{pos}$, and (i) the correlation between $\nabla \times B_{pos}$ and $n_{H_2}$ for the velocity channel at $+3.090$ km s$^{-1}$, where $x$ is the axis of right ascension from west to east and $y$ is the axis of declination from south to north.
Segments in the $n_{H_2}$ and $\delta \phi$ maps show the polarization segments in Fig.\ \ref{fig1}.
Vectors in the maps of $B_{pos}$, $B_x$, and $B_y$ show the magnetic field direction with length proportional to the field strength. 
The black boxes in the maps of $n_{H_2}$, $\delta \phi$, $B_{pos}$, $B_x$, $B_y$, $\partial B_x$/$\partial y$, and $\partial B_y$/$\partial x$ show the pixels with $\nabla \times B_{pos}$ data points.}
\end{figure*}

\renewcommand{\thefigure}{Figure 4}
\begin{figure*}
\centering
\includegraphics[width=1.\textwidth]{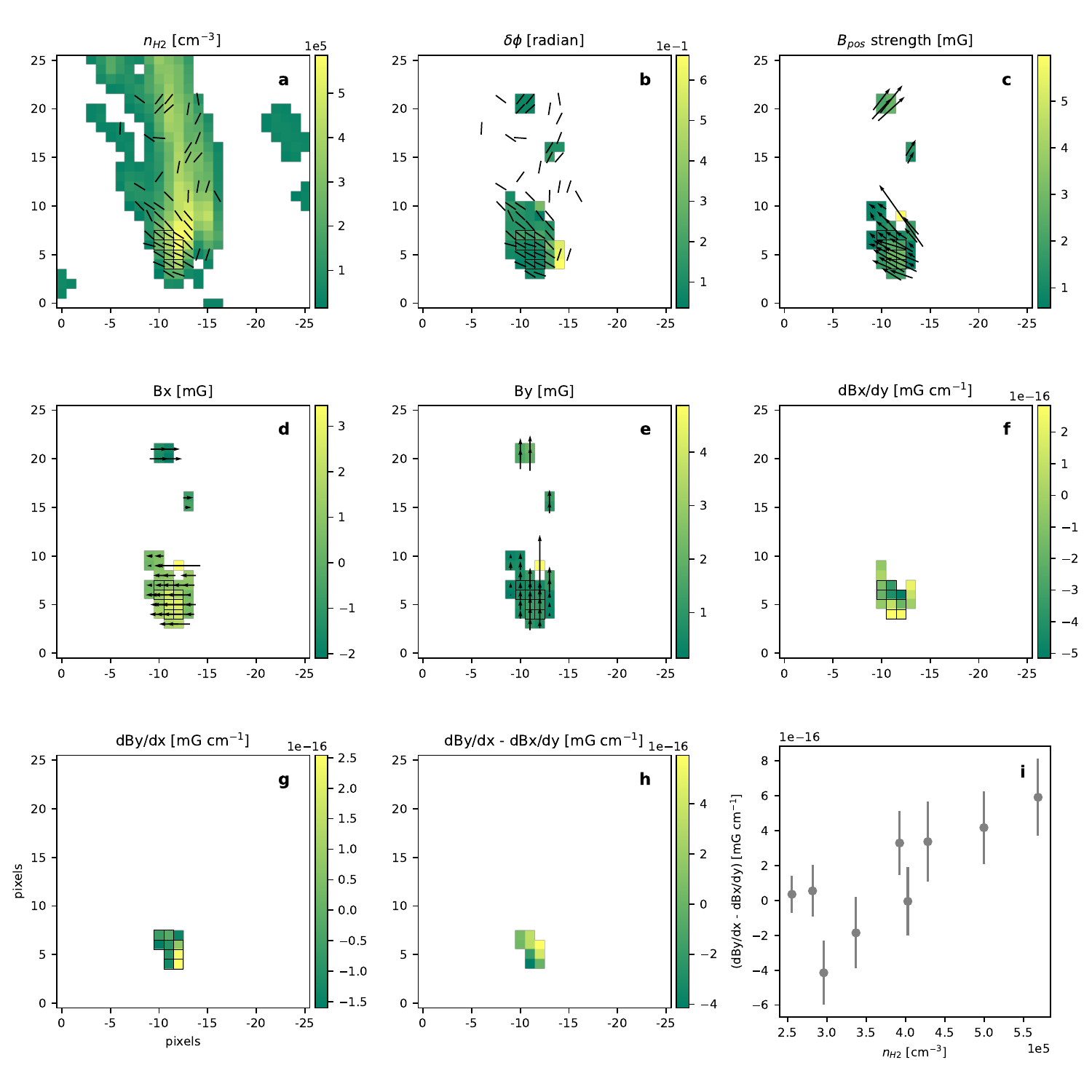}
\caption{The same as Extended Data Figure 3 but for the velocity channel at $+6.265$ km s$^{-1}$.
The directions of the $B_{pos}$, $B_x$, and $B_y$ vectors in this plot are opposite to those in Extended Data Figure 3 because the velocity channels at $+6.265$ and $+3.090$ km s$^{-1}$ trace the background (the red box in Extended Data Figure 2) and foreground (the blue box in Extended Data Figure 2) layers of the outflow cavity, respectively.}
\end{figure*}

\renewcommand{\thefigure}{Figure 5}
\begin{figure*}
\centering
\includegraphics[width=1.\textwidth]{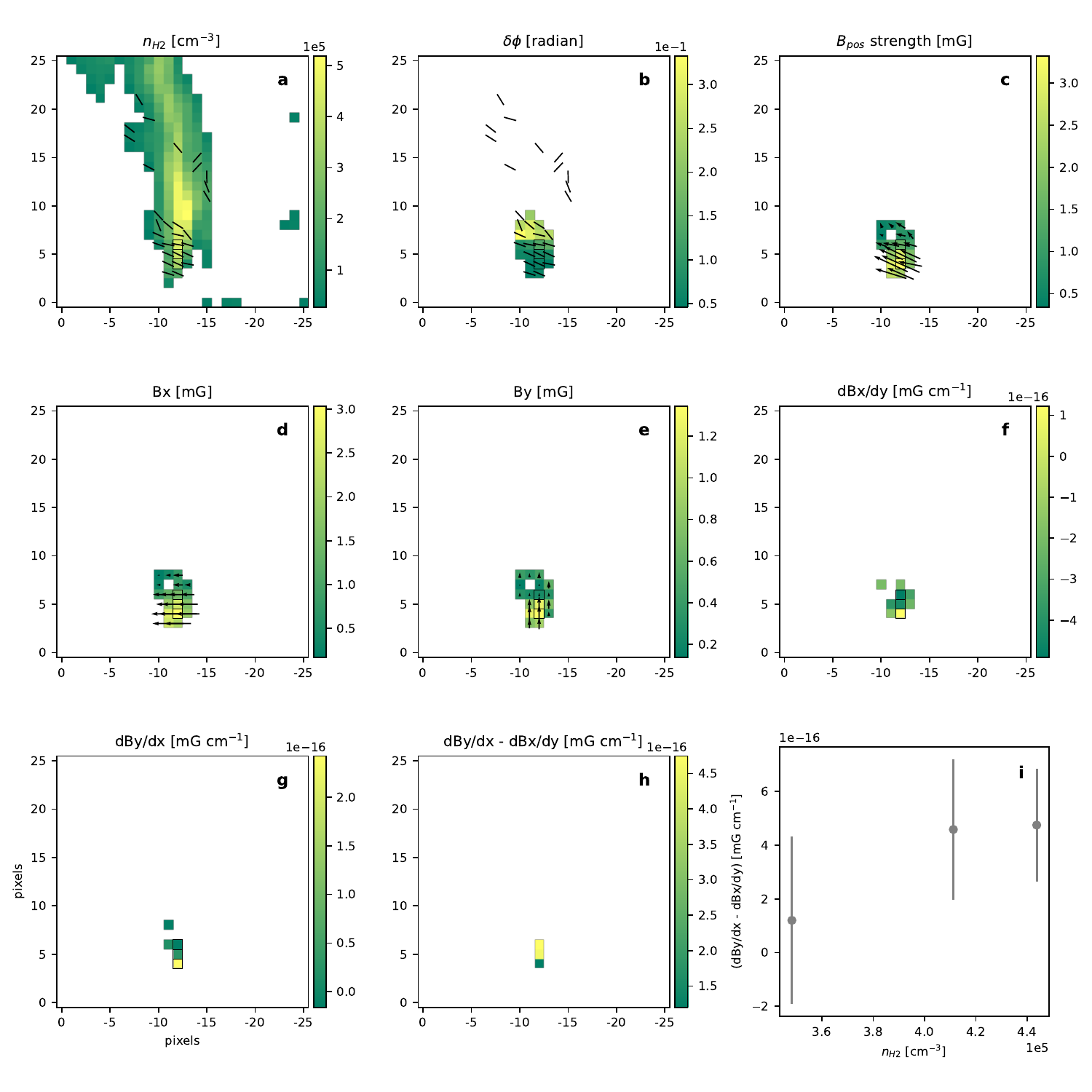}
\caption{The same as Extended Data Figure 3 but for the velocity channel at $+9.440$ km s$^{-1}$.
The directions of the $B_{pos}$, $B_x$, and $B_y$ vectors in this plot are opposite to those in Extended Data Figure 3 because the velocity channels at $+9.440$ and $+3.090$ km s$^{-1}$ trace the background (the red box in Extended Data Figure 2) and foreground (the blue box in Extended Data Figure 2) layers of the outflow cavity, respectively.}
\end{figure*}

\renewcommand{\thefigure}{Figure 6}
\begin{figure*}
\centering
\includegraphics[width=1.\textwidth]{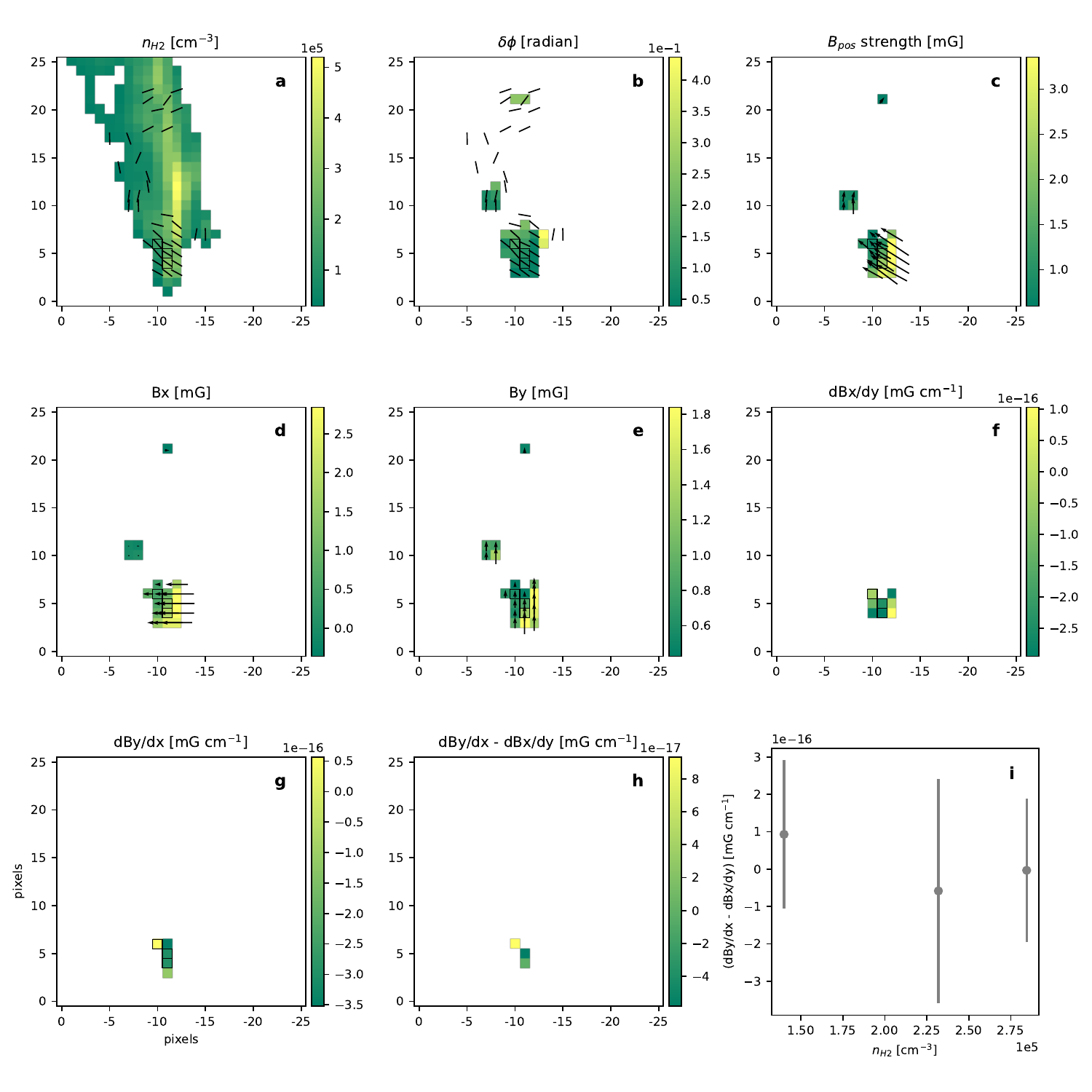}
\caption{The same as Extended Data Figure 3 but for the velocity channel at $+12.615$ km s$^{-1}$.
The directions of the $B_{pos}$, $B_x$, and $B_y$ vectors this plot are opposite to those in Extended Data Figure 3 because the velocity channels at $+12.615$ and $+3.090$ km s$^{-1}$ trace the background (the red box in Extended Data Figure 2) and foreground (the blue box in Extended Data Figure 2) layers of the outflow cavity, respectively.}
\end{figure*}

\renewcommand{\thefigure}{Figure 7}
\begin{figure*}
\centering
\includegraphics[width=1.\textwidth]{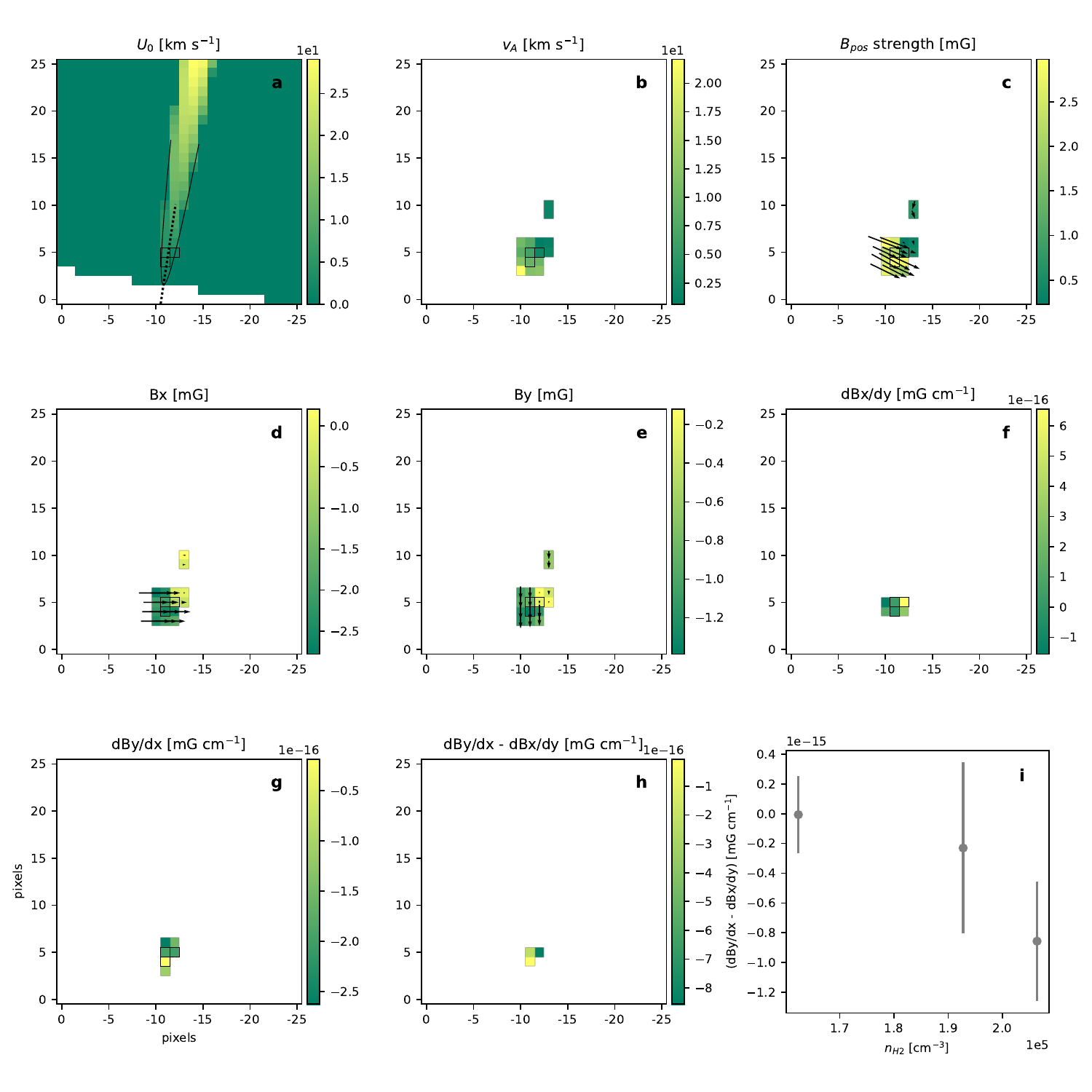}
\caption{{The same as Extended Data Figure 3, but showing the derivation of $B_{pos,mDCF}$. The $n_{H_2}$ and $\delta \phi$ are identical to those used in Extended Data Figure 3, and here panels \textbf{a} and \textbf{b} show the $U_0$ and $v_A$, respectively. The dotted line and parabolic curve in panel \textbf{a} show the axis of the 4A1 outflow and the shell of the high-velocity outflow.}}
\end{figure*}

\renewcommand{\thefigure}{Figure 8}
\begin{figure*}
\centering
\includegraphics[width=1.\textwidth]{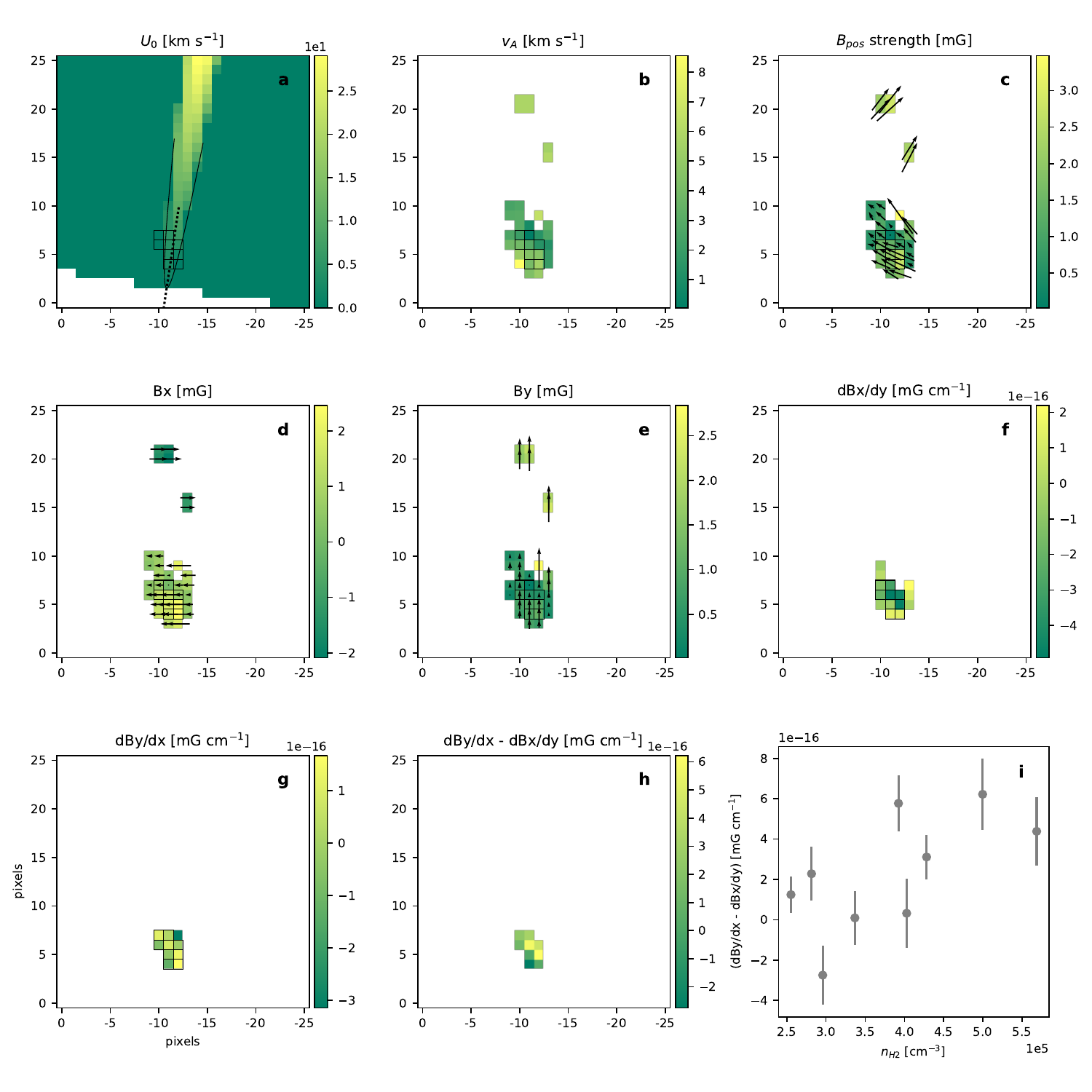}
\caption{{The same as Extended Data Figure 7 but for the velocity channel at +6.265 km s$^{-1}$}}
\end{figure*}

\renewcommand{\thefigure}{Figure 9}
\begin{figure*}
\centering
\includegraphics[width=1.\textwidth]{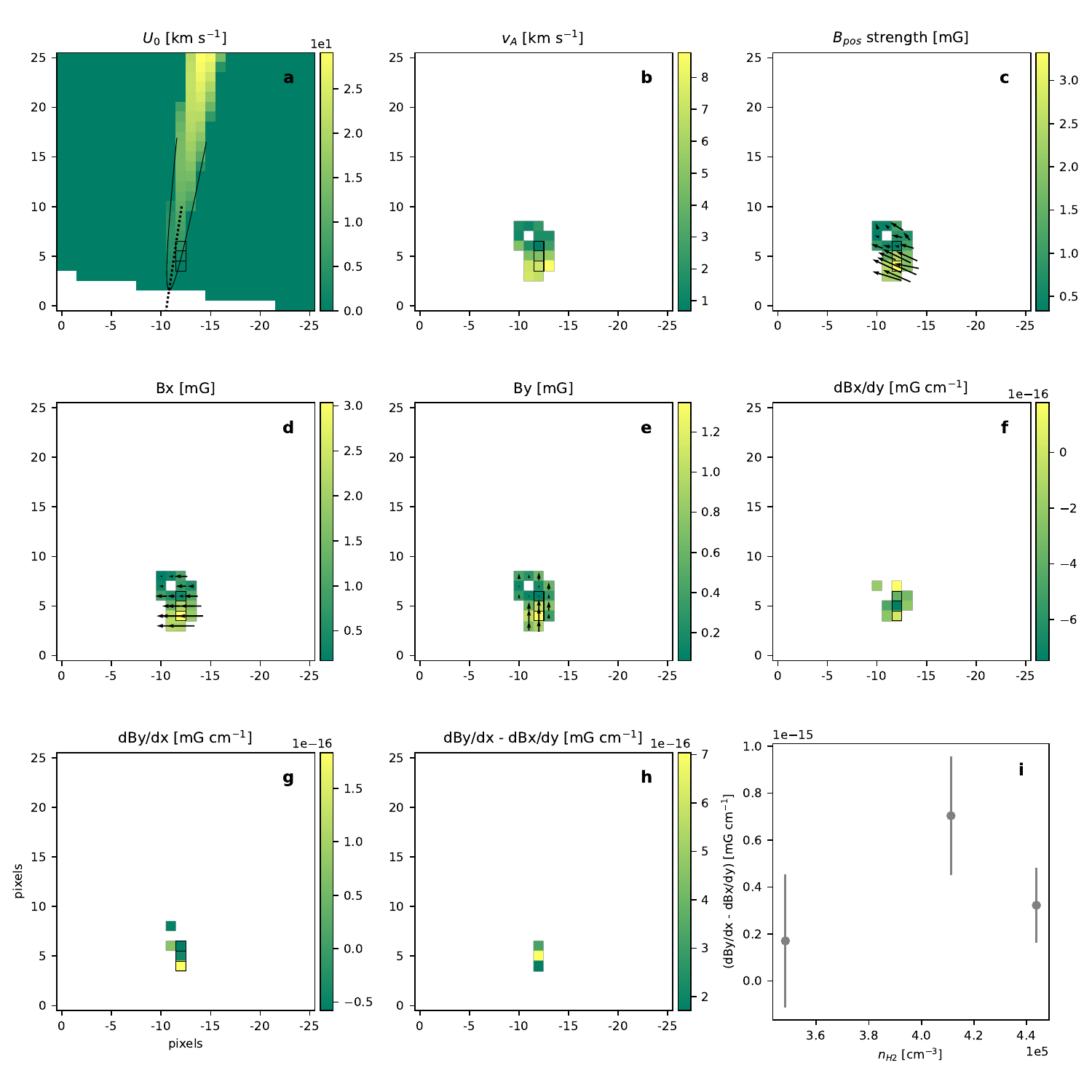}
\caption{{The same as Extended Data Figure 7 but for the velocity channel at +9.440 km s$^{-1}$}}
\end{figure*}

\renewcommand{\thefigure}{Figure 10}
\begin{figure*}
\centering
\includegraphics[width=1.\textwidth]{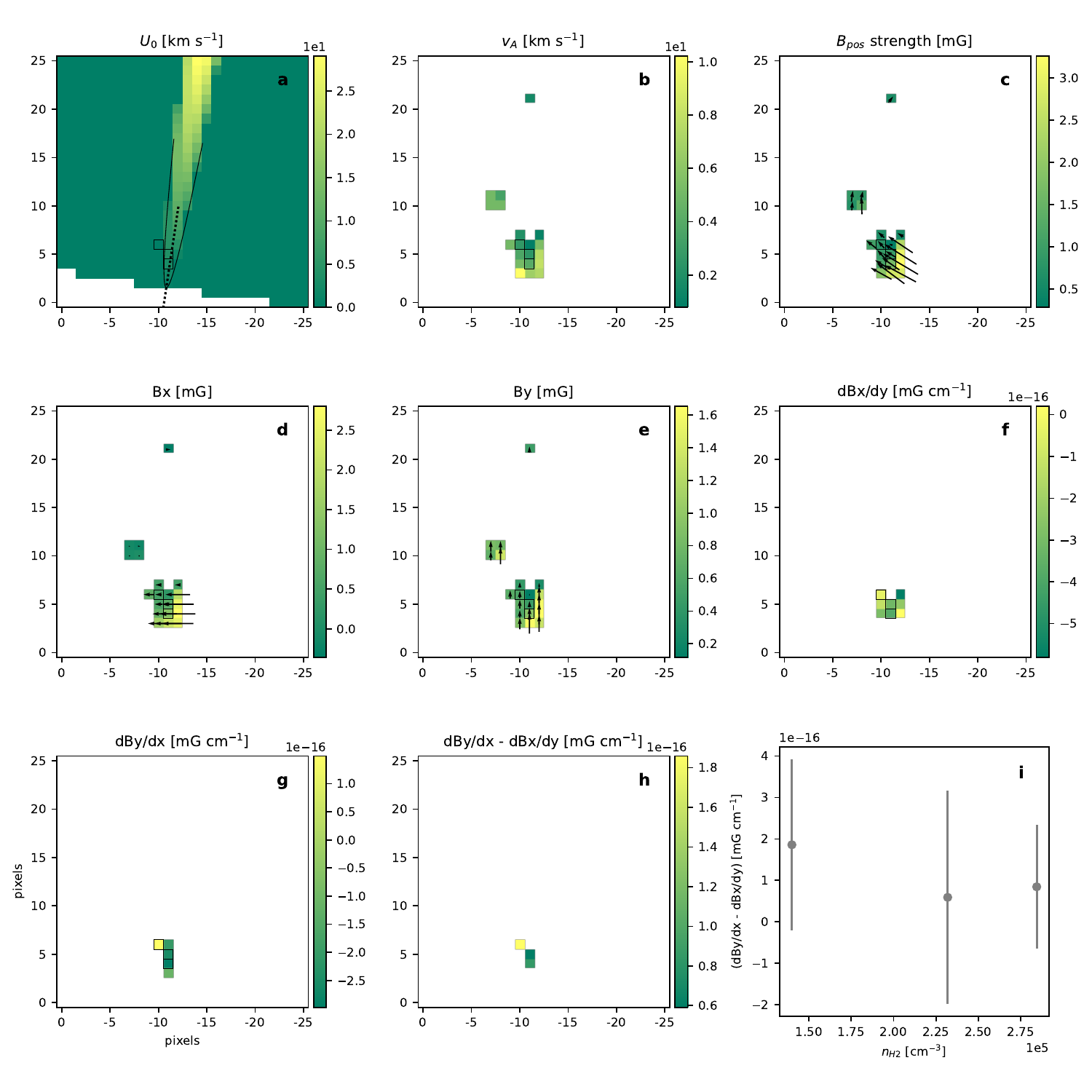}
\caption{{The same as Extended Data Figure 7 but for the velocity channel at +12.615 km s$^{-1}$}}
\end{figure*}

\renewcommand{\thefigure}{Figure 11}
\begin{figure*}
\centering
\includegraphics[width=1.\textwidth]{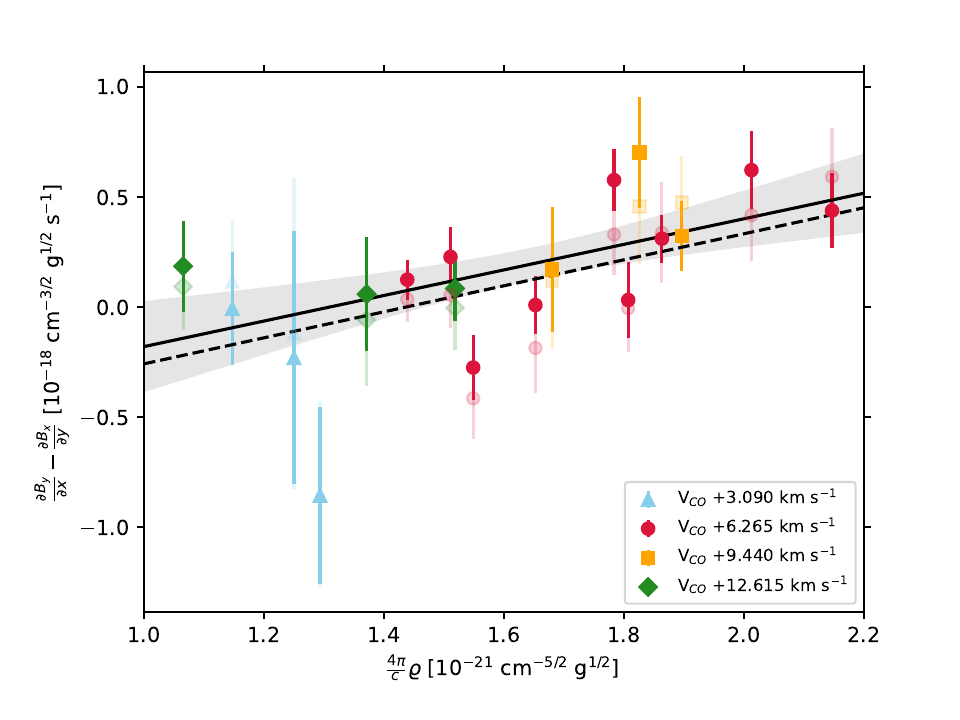}
\caption{{The same analysis as Fig.\ \ref{fig3} with $B_{pos,mDCF}$ substituted for $B_{pos}$. The data points in faint colors and the dashed line represent the data points and linear regression in Fig.\ \ref{fig3}.}}
\end{figure*}

\renewcommand{\thefigure}{Figure 12}
\begin{figure*}
\centering
\includegraphics[width=1.\textwidth]{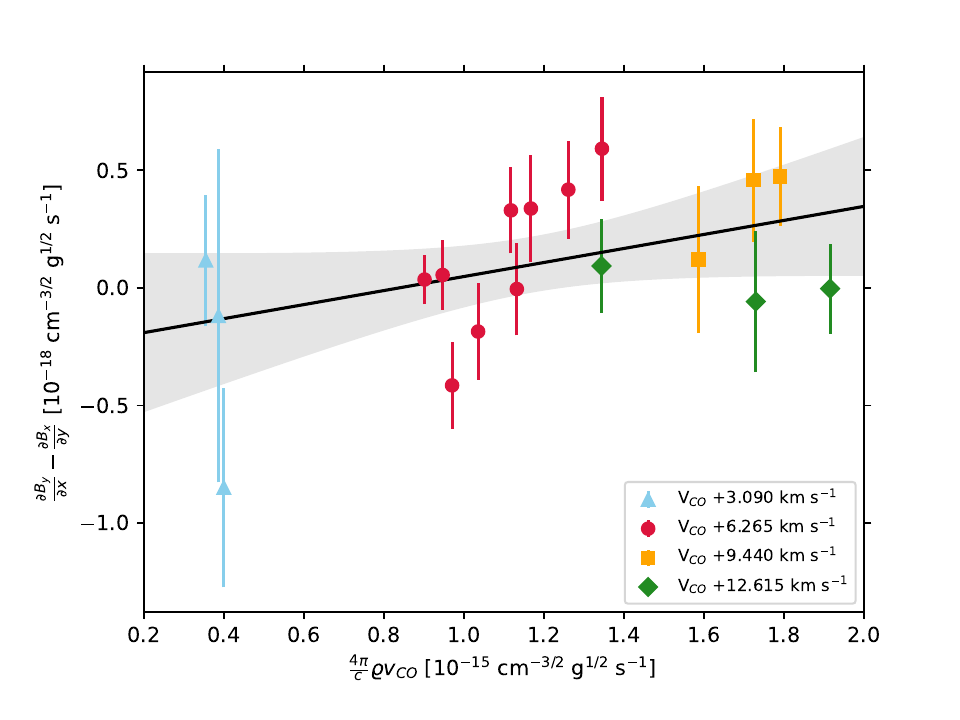}
\caption{The same as Fig.\ \ref{fig3} but assuming that $v_d$ is proportional to the outflow velocity $v_{CO}$.}
\end{figure*}
\clearpage

\renewcommand{\thefigure}{Figure 13}
\begin{figure*}
\centering
\includegraphics[width=0.7\textwidth]{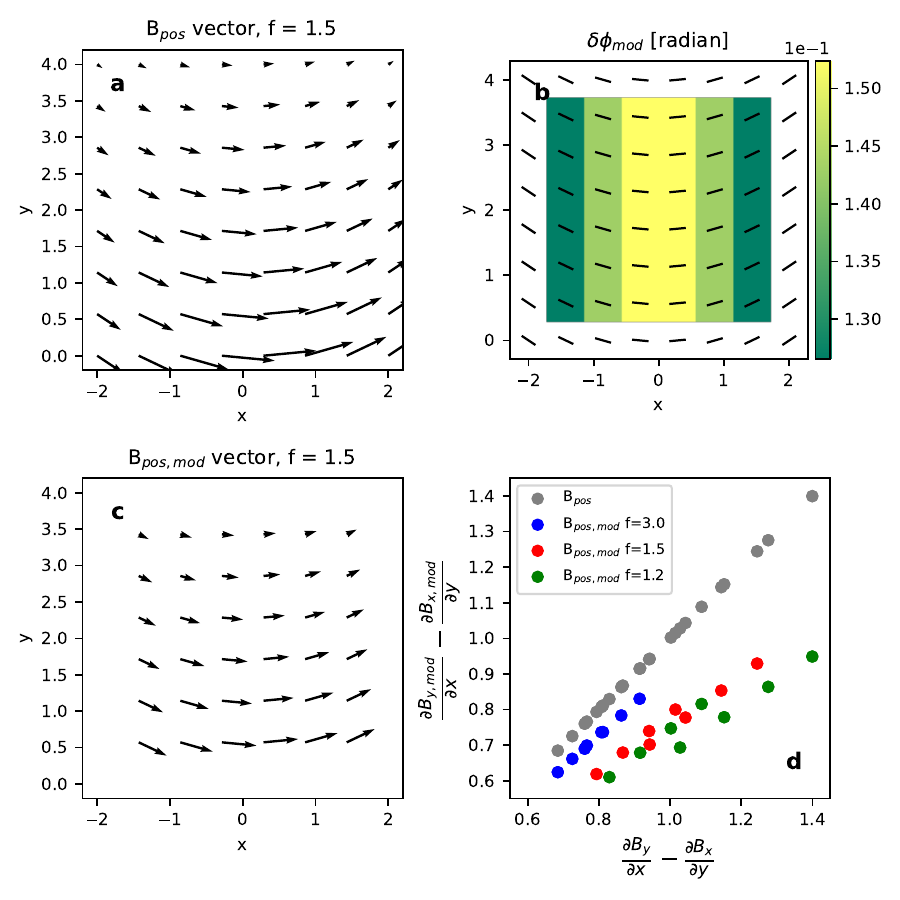}
\caption{{Simulation of the unsharp masking method on modifying $\frac{\partial B_y}{\partial x}-\frac{\partial B_x}{\partial y}$. 
\textbf{a}, input parabolic magnetic fields $B_{pos}$ with focal length $f = 1.5$. Vectors show the magnetic field direction with lengths proportional to field strength in arbitrary units. \textbf{b}, $\delta \phi_{mod}$ derived from the dispersion of magnetic field position angles of the input model within 3 $\times$ 3 pixels in color scale. Black segments represent the polarization orientations of the input model. \textbf{c}, modified magnetic fields $B_{pos,mod}$ after taking $\delta \phi_{mod}$ into account. The ratio between vector length and field strength is the same as panel \textbf{a}. \textbf{d}, the correlation between $\nabla \times B_{pos}$ and $\nabla \times B_{pos,mod}$. The gray dots show the initial data. The blue, red, and green dots show the data modified by the unsharp masking method for the models of $f = 3, 1.5, 1.2$, respectively.}}
\end{figure*}
\clearpage

\renewcommand{\thefigure}{Figure 14}
\begin{figure*}
\centering
\includegraphics[width=1.\textwidth]{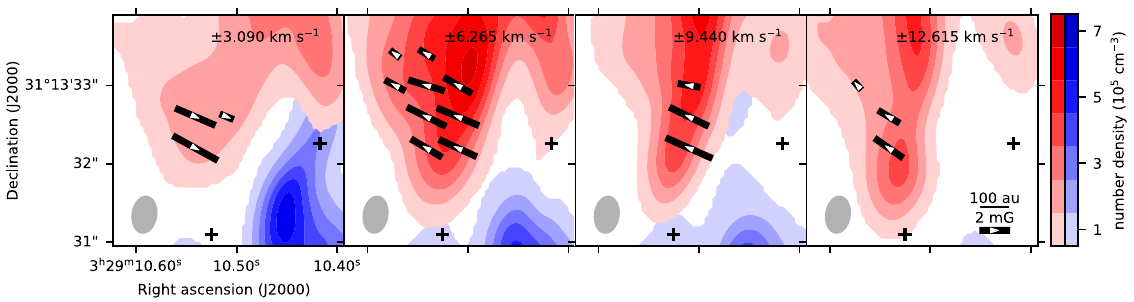}
\caption{The same as Fig.\ \ref{fig4} but showing the $B_{pos}$ direction if the line-of-sight current of the 4A1 redshifted outflow is pointing away from us.}
\end{figure*}

\renewcommand{\thefigure}{Figure 15}
\begin{figure*}
\centering
\includegraphics[width=1.\textwidth]{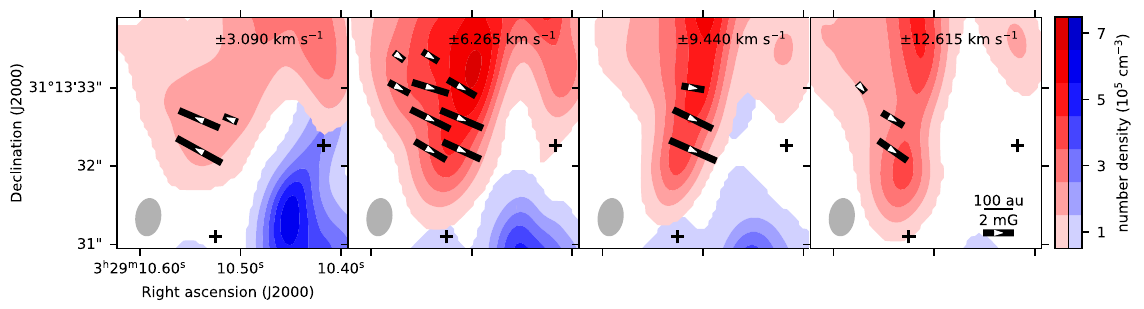}
\caption{The same as Fig.\ \ref{fig4} but showing the $B_{pos}$ direction if the line-of-sight current of the 4A1 redshifted outflow is pointing toward us.}
\end{figure*}

\renewcommand{\thefigure}{Figure 16}
\begin{figure*}
\centering
\includegraphics[width=1.\textwidth]{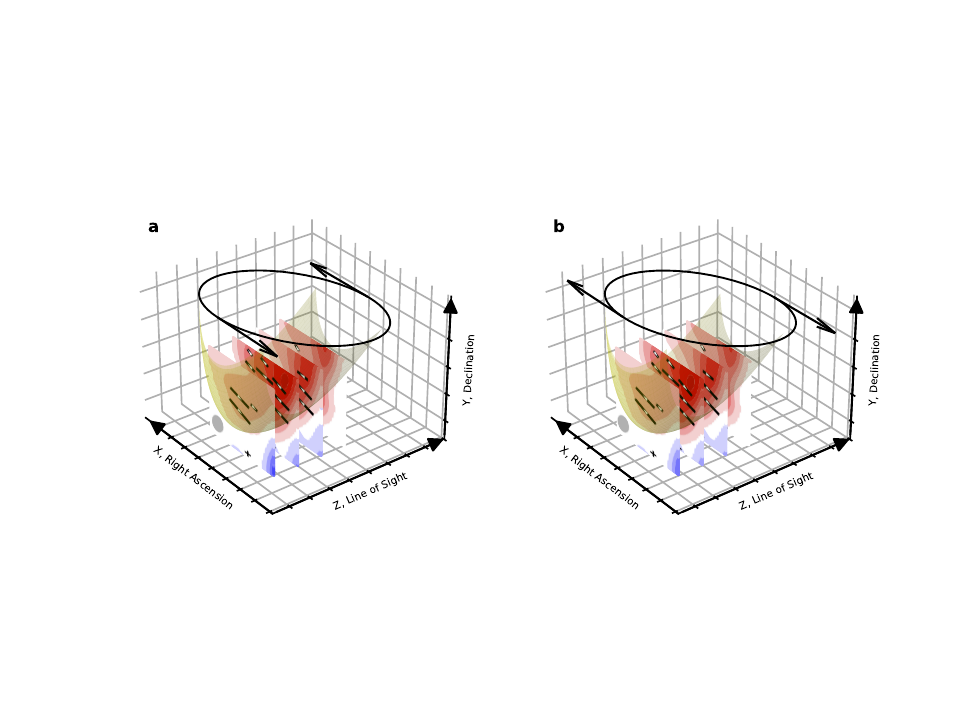}
\caption{Illustrations of the 3-dimensional geometry of the 4A1 outflow and magnetic fields. 
{The arrows of the X, Y, Z axes represent the coordinate system adopted in this work.}
\textbf{a}, the outflow with line-of-sight current pointing away from us. The parabolic corn represents the outflow cavity, and the black circle and arrows show the toroidal magnetic field counter-clockwisely wrapping around the outflow. In this case, the magnetic field direction in the foreground CO channel map is in the southwest direction.  \textbf{b}, the outflow with line-of-sight current pointing toward us. In this case, the magnetic field direction in the foreground CO channel map is in the northeast direction.}
\end{figure*}

\end{document}